\documentclass[iop]{emulateapj}

\usepackage[]{graphicx, amsmath, natbib, multirow}

\slugcomment{Accepted by The Astrophysical Journal, August 19, 2013}

\shorttitle{PALM-3000 Instrument Paper}
\shortauthors{Dekany et al.}

\begin{document}

\title{PALM-3000: Exoplanet adaptive optics for the 5-meter Hale Telescope}

\author{Richard Dekany\altaffilmark{1}, Jennifer Roberts\altaffilmark{2}, Rick Burruss\altaffilmark{2}, Antonin Bouchez\altaffilmark{1,3}, Tuan Truong\altaffilmark{2}, Christoph Baranec\altaffilmark{1}, Stephen Guiwits\altaffilmark{2}, David Hale\altaffilmark{1}, John Angione\altaffilmark{2}, Thang Trinh\altaffilmark{2}, Jeffry~Zolkower\altaffilmark{1,4}, J.~Christopher~Shelton\altaffilmark{2}, Dean Palmer\altaffilmark{2}, John Henning\altaffilmark{1}, Ernest Croner\altaffilmark{1}, Mitchell~Troy\altaffilmark{2}, Dan~McKenna\altaffilmark{1}, Jonathan~Tesch\altaffilmark{2},~Sergi~Hildebrandt\altaffilmark{1,2},~Jennifer~Milburn\altaffilmark{1}}

\altaffiltext{1}{Caltech Optical Observatories, California Institute of Technology, 1200 E. California Blvd., MC 11-17, Pasadena, CA 91125, USA; rgd@astro.caltech.edu}
\altaffiltext{2}{Jet Propulsion Laboratory, California Institute of Technology, 4800 Oak Grove Blvd., Pasadena, CA  91109, USA}
% \altaffiltext{3}{Control and Dynamical Systems, California Institute of Technology, 1200 E. California Blvd., MC 107-81, Pasadena, CA 91125}
% \altaffiltext{4}{Department of Astronomy, California Institute of Technology, 1200 E. California Blvd., MC 249-17, Pasadena, CA 91125}
\altaffiltext{3}{Now at Giant Magellan Telescope Observatory Corp., Pasadena, CA  91106, USA}
\altaffiltext{4}{Now at Department of Astronomy, Cornell University, Ithaca, NY 14853, USA}

% Notice that each of these authors has alternate affiliations, which
% are identified by the \altaffilmark after each name. Specify alternate
% affiliation information with \altaffiltext, with one command per each
% affiliation.
    
\begin{abstract}

We describe and report first results from PALM-3000, the second-generation astronomical adaptive optics facility for the 5.1-m Hale telescope at Palomar Observatory. PALM-3000 has been engineered for high-contrast imaging and emission spectroscopy of brown dwarfs and large planetary mass bodies at near-infrared wavelengths around bright stars, but also supports general natural guide star use to V $\approx$ 17. Using its unique 66 $\times$ 66 actuator deformable mirror, PALM-3000 has thus far demonstrated residual wavefront errors of 141~nm RMS under $\sim$1\arcsec~seeing conditions. PALM-3000 can provide phase conjugation correction over a 6\farcs4 $\times$ 6\farcs4 working region at $\lambda = 2.2$ $\mu$m, or full electric field (amplitude and phase) correction over approximately one half of this field.  With optimized back-end instrumentation, PALM-3000 is designed to enable $10^{-7}$ contrast at 1\arcsec~angular separation, including post-observation speckle suppression processing. While optimization of the adaptive optics system is ongoing, we have already successfully commissioned five back-end instruments and begun a major exoplanet characterization survey, Project 1640, with our partners at American Museum of Natural History and Jet Propulsion Laboratory.

\end{abstract}

% Keywords should appear after the \end{abstract} command. The uncommented
% example has been keyed in ApJ style. See the instructions to authors
% for the journal to which you are submitting your paper to determine
% what keyword punctuation is appropriate.


% From the front matter, we move on to the body of the paper.
% In the first two sections, notice the use of the natbib \citep
% and \citet commands to identify citations. The citations are
% tied to the reference list via symbolic KEYs. The KEY corresponds
% to the KEY in the \bibitem in the reference list below. We have
% chosen the first three characters of the first author's name plus
% the last two numeral of the year of publication as our KEY for
% each reference.

\section{Introduction}
Adaptive optics (AO) systems have proven essential in opening key new areas of astronomical research by compensating for atmospheric blurring to achieve diffraction-limited infrared imaging and spectroscopy.  The successful development of AO capabilities on the world's large-aperture telescopes over the past two decades is a scientific achievement of a large and dedicated instrumentation community \citep{Hart:10, 2012ARA&A..50..305D}.  The Palomar Adaptive Optics System (PALM-241) on the 5.1-m Hale telescope at Palomar Mountain ~\citep{1997SPIE.3126..269D, 2000SPIE.4007...31T, 2006SPIE.6272E..13D} has been a leader in astronomical AO, providing facility-class on-sky operation since December 1999.  The instrument, jointly-developed by Jet Propulsion Laboratory and Caltech, was used to obtain data contributing to $\approx 100$ refereed astronomical journal articles while facilitating extensive AO technology development.  The advantages of accessible site, moderate altitude, and excellent engineering support infrastructure have supported the on-sky deployment of many innovations: the first high-speed AO telemetry recording system~\citep{2003SPIE.4839..911T}, the now widespread SciMeasure high-speed wavefront sensor camera~\citep{2000ASSL..252..395D, 2000SPIE.4007..481D, 2001ExA....11..237D}, the first high-order natural guide star (NGS) Shack-Hartmann simultaneous multiple star wavefront sensor~\citep{2006SPIE.6272E.169V}, demonstrations of sparse matrix~\citep{2003SPIE.4839.1035S}, hierarchical~\citep{MacMartin:03}, efficient recursive ~\citep{Ren:05} and Fourier transform based slope sensor reconstructors~\citep{Poyneer:03}, AO-assisted `lucky' imaging at visible wavelengths~\citep{2009ApJ...692..924L}, vector-vortex coronagraph~\citep{Serabyn:09}, non-redundant mask interferometry~\citep{2010ApJ...715..724B}, and three years of solid-state sum-frequency 589~nm laser guide star operations~\citep{2004SPIE.5490.1033V,2007AAS...211.5803B,2008SPIE.7015E..74R} including the demonstration of back pumping and frequency chirping techniques that increase the efficiency of photoreturn from Earth's mesospheric sodium layer~\citep{2008SPIE.7015E..14K, 2009amos.confE..33K}.  PALM-241 achieved infrared K-band imaging Strehl ratios under median seeing conditions of 1\farcs1 as high as 73\% as measured using point spread function fitting photometry  \citep{2004SPIE.5490..504R}.  Exposures of at least 30 seconds are needed with our instrumentation to estimate infrared Strehl ratios properly, accounting for the faint, extended contribution of the point spread function halo.  Assuming Marechal's approximation \citep{Marechal47}, this is an equivalent RMS wavefront error of 195~nm. Using a centrally-projected guide star laser emitting up to 8 W of 589~nm sodium $D_2$ light to form a mesospheric synthetic beacon, PALM-241 achieved 48\% K-band imaging Strehl as part of a laser proof-of-concept demonstration science program \citep{2008SPIE.7015E..74R}.

Since the initial discovery of a confirmed substellar object at Palomar Observatory \citep{Nakajima:95}, low-mass objects have been a focus of direct imaging searches and have motivated the Palomar AO program.  Despite excellent performance and successful investigations of brown dwarf companions \citep{2003A&A...410..283B, 2006ApJ...650L.131L, 2006A&A...460..799G, 2009ApJS..181...62M}, the ability of PALM-241 to directly image stellar companion exoplanets using the full 5.1-m aperture was limited to a contrast of $\sim$$10^{-4}$ at 0\farcs5 by imperfect complex amplitude correction of the electromagnetic field at the coronagraphic occulting mask, resulting in residual semi-static stellar speckles \citep{2000SPIE.4007..889B, 2000SPIE.4007..899O, 2001ApJ...558L..71B}.  These speckles would vary in amplitude as well as contribute photon noise in the focal plane.  Improving exoplanet imaging capabilities beyond these limitations, to approach the theoretical limits for ground-based exoplanet study \citep{1994Natur.368..203A}, required a combination of improved calibration, post-processing techniques, and capability of the AO system.

PALM-3000 is the second-generation astronomical AO system for the Hale telescope.  It has been engineered specifically to optimize high-contrast studies of exoplanets orbiting nearby bright stars when used in conjunction with specialized speckle-suppressing instrumentation, including P1640 \citep{2011PASP..123...74H, 2011PASP..123..746Z}, the Palomar High-Angular Resolution Observer (PHARO) \citep{2001PASP..113..105H},  and the Palomar Fiber Nuller (PFN) \citep{2011ApJ...729..110H}.  The combination of PALM-3000 and P1640, in particular, is expected to reach $5\sigma$ detection contrast ratios as high as $10^{-7}$ at an angular radius of 1\farcs0\ and $10^{-4}$ at 0\farcs3 after application of speckle suppression algorithms exploiting the coherent integral field spectroscopic data cube available with P1640.

To date, our team has conducted commissioning observations with PALM-3000 and five visible and near-infrared science instruments.  In particular, we have quantified the benefit of the speckle suppression technique using P1640, in terms of the faintest companion that can be detected with $5 \sigma$ confidence as a function of separation from a star.  Initial observations achieved a contrast of $10^{-5}$ at 0\farcs5 angular separation from an occulted star \citep{2012SPIE.8447E..20O}, suitable for detecting luminous sub-stellar companions, though our integrated system is not yet at full performance awaiting improved instrument calibration and AO performance improvements.  Future studies will report on the results of our optimization activities.

\section{Science Motivation}
\subsection{Exoplanet Spectroscopy}
The direct imaging discovery of planetary mass companions to HR 8799 \citep{2008Sci...322.1348M}  and Fomalhaut \citep{2008Sci...322.1345K} with orbital radii of tens of AU has inaugurated a new era in the study of exoplanets.  Radial velocity and transit detection techniques, so fruitful for exoplanet discovery, are currently only sensitive to detection of a strongly-irradiated class of planets orbiting close in to their host stars.  Fully mature high-contrast direct imaging campaigns, on the other hand, will be sensitive to companions at large separations and will probe a critical region of the exoplanetary parameter space.  Direct imaging studies will allow us to address questions about solar system architectures by evaluating the frequency of companions undetectable by radial velocity and transit methods.  Even a flat extrapolation of the semi-major axis distribution of radial-velocity detected exoplanets beyond 3-5 AU foreshadows a large population of undetected planets around nearby stars \citep{2011ApJ...733..126C}.  In addition, the increasing frequency of planet masses discovered  through radial velocity techniques below about 4 $M_{Jup}$ suggests a large population of lower mass planets exist around nearby stars.  Moreover, high-contrast imaging campaigns will further constrain our understanding of the distribution of brown dwarf masses and their frequency around a variety of host stars.  High-contrast integral field spectrographs (IFS) will allow study of massive exoplanets in detail, tackling questions about their internal physics, structure, and atmospheric chemistry.

A survey has been designed \citep{2011PASP..123...74H} to search for and characterize exoplanets during 99 nights of dedicated observing with PALM-3000 and P1640.  Motivated by the correlation between planet frequency and the mass of the host star \citep{2007ApJ...670..833J}, the survey will target approximately 220 stellar systems having A and F-type stars, a population expected to host exoplanets 5 times as often as their lower mass counterparts.  Indeed, two of the first imaged exoplanet systems, HR 8799 and Fomalhaut, have A-type hosts. A secondary survey will also be performed of  stars within 25 pc with visual magnitude V $<$ 7 establishing a complete census of wide-orbit candidates.  Already, PALM-3000 and the P1640 instrument have obtained simultaneous infrared spectra of the HR8799 giant planets \citep{2013ApJ...768...24O}, moving exoplanet study into a spectrographic science.  With the expected final adaptive optics performance of PALM-3000 (\S\ref{sec:predicted_performance}) we will be able to observe upwards of 90\% of the planets of greater than 7 Jupiter masses orbiting the stars in our sample \citep{2011PASP..123...74H, 2010PASP..122..162B}.  This survey will comprehensively determine whether the theorized brown dwarf 'desert' \citep{2008ApJ...679..762K} exists.  Depending on the true fraction of stars harboring exoplanets, we will additionally obtain low-resolution spectra of several planet-mass objects in each of several distinctly different age bins.  

\subsection{Circumstellar Disks}
A comprehensive picture of the formation of planets can only be obtained with a thorough understanding of the structure and evolution of the circumstellar disks from which they evolve.  After 5-10 Myr, most of the gas content of the primordial, massive and optically-thick circumstellar disk has been cleared away \citep{2001ApJ...553L.153H}, accreted onto the star itself, formed into planetary bodies, or dispersed via stellar radiation.  Stars exiting this primordial phase possess a much less massive optically-thin disk, mostly containing second-generation grains, formed by the collisional fragmentation of larger bodies.  IRAS and Spitzer have detected hundreds of debris disks by their far-infrared excesses \citep{2008ARA&A..46..339W}, but only a dozen images of optically thin disks are currently available (see circumstellardisks.org for an up-to-date list).  Multi-wavelength resolved imaging is a necessary complement to the infrared spectral energy distribution analysis to remove degeneracies between grain morphology, temperature distribution, and grain size.  The range and values of these parameters frame the conditions for planet formation.

Furthermore, the most prominent disk structures (e.g. warp, annulus, asymmetries) match the predicted dynamical effects based on the characteristics of embedded planets \citep{1999ApJ...527..918W, 2009ApJ...693..734C, 2012A&A...542A..41C}.  The Hubble Space Telescope has been a prime observatory for high-contrast imaging at short wavelengths ($\lambda = 0.5-2.4$ $\mu$m) for a decade, thanks to the ACS and NICMOS coronagraphs despite the large radii of their coronagraphic obscurations, i.e. large inner working angle (IWA), 0\farcs9 for ACS, 0\farcs5 for NICMOS.  Unfortunately these two exceptional instruments are no longer in service.  The James Webb Telescope will provide only modest short wavelength capabilities, as it will not be diffraction limited in the short near-IR and red visible bands where PALM-3000 will perform best. 

Using both P1640 (simultaneously recording both J and H band spectra) and PHARO (providing K band imagery), PALM-3000 will look across $\lambda = 1.2-2.5$ $\mu$m at known and yet-to-be identified spatially resolved debris disks at small IWA's, where warmer dust is likely to reside and where rocky planet formation is likely to occur (1-10 AU).  Imaging such spatial scales is necessary to understand the interaction between the debris disk and planetary bodies, as well as the implications for planetary formation mechanisms.  PALM-3000 will also characterize bright new candidates from the latest Spitzer and Herschel surveys.  The sensitivity in H and K bands will be background-limited to 17-16 mag/arcsec$^2$, respectively, which will allow imaging of brighter circumstellar disks around young stars, those having a ratio of infrared luminosity relative to their host star $(LIR/L*) > 10^{-3}$ at high signal-to-noise ratio.

\subsection{Well-corrected subaperture mode inside 200 milli-arc seconds }
PALM-3000 retains the unique off-axis subaperture mode of PALM-241 \citep{Serabyn:09}. When used with the visible fast-frame imaging camera TMAS (Table~\ref{tab:Instruments}) and a vector-vortex coronagraphic mask \citep{doi:10.1117/12.858240}, it will be possible to obtain high-contrast visible-light images of circumstellar material for guide stars of V $<$ 2.5.  This high-contrast parameter space at extremely small IWA is largely unexplored \citep{2012SPIE.8442E..04M, 2013A&A...552L..13M} and may include the possibility of detecting the first exoplanet in reflected light \citep{doi:10.1117/12.894309}.  Simulations of this mode using the wave-optics propagation software PAOLA \citep{2006JOSAA..23..382J}  indicate an R-band Strehl ratio of $>$ 90\% is possible in good seeing conditions at Palomar, far exceeding reported visible-light coronagraphic rejection to date \citep{Swartzlander08}.  Because the Monte-Carlo based models of the frequency and distribution of detectable reflection exoplanets indicates the total number of detectable reflection exoplanets will be very limited \citep{2011PASP...McBride}, this mode is best considered a technique for targeted studies of known objects identified through radial velocity or transit photometry discovery.  Similarly, contrast of better than $10^{-5}$ at an angular separation less than 0\farcs2 has exciting possibilities for new studies of stellar astrophysics such as direct imaging of Herbig-Halo outflows \citep{2000SPIE.4007..839B, doi:10.1117/12.857245}, Mira stars \citep{2011MNRAS.417...32L}, and supergiant environments \citep{2009A&A...508..923H}. 

\section{System Description}

\subsection{Optical Design}\label{OpticalDesign}
\begin{figure*}[h!]
\center
\includegraphics[angle=90,width=5.5in]{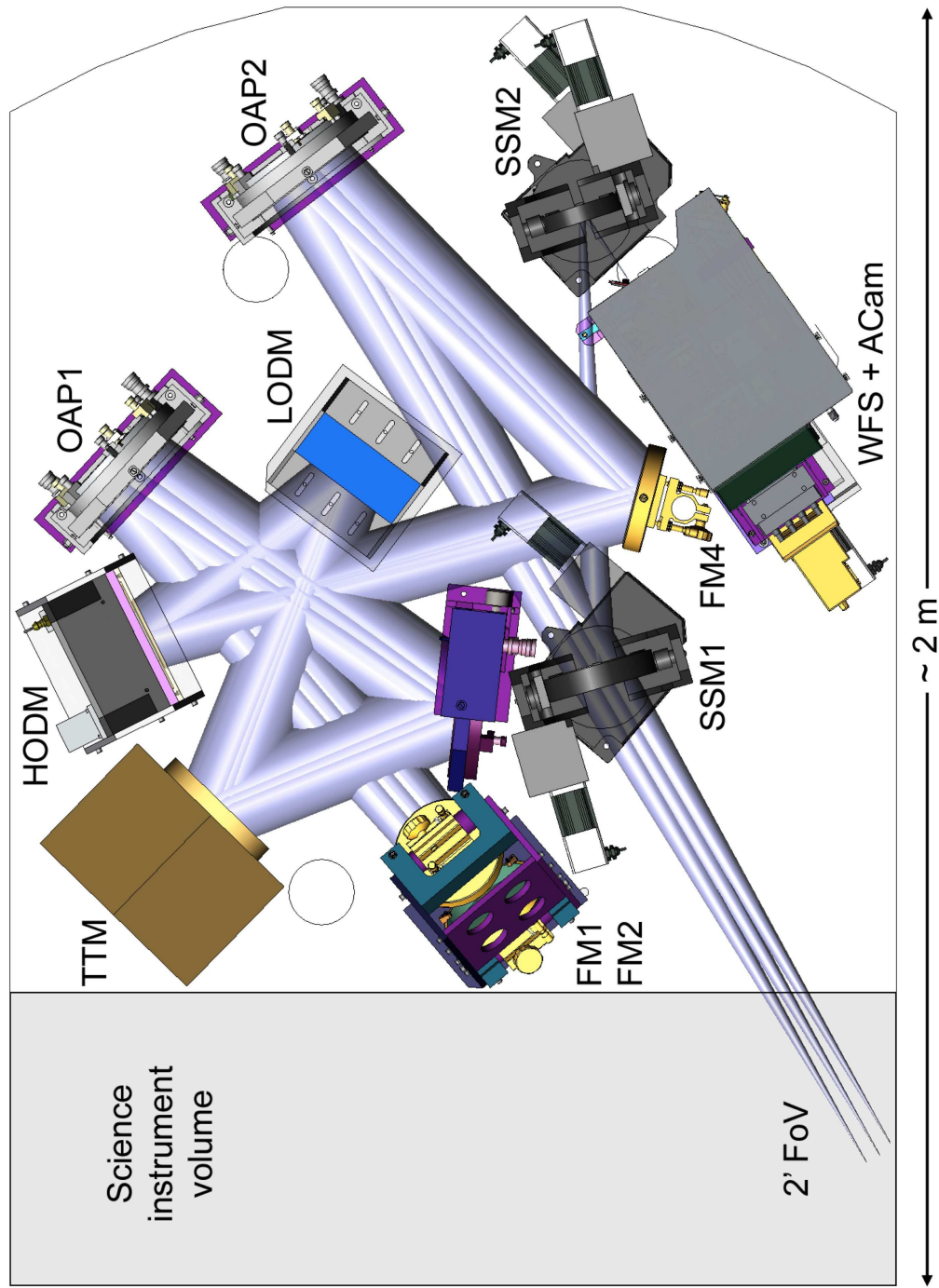}
\caption{ PALM-3000 optical layout, viewed from beneath the Cassegrain cage.  Light passing through the PALM-3000 optical bench is folded by path-reducing fold mirrors, FM1 and FM2, to the collimating parabola, OAP1.  In collimated space, the beam is folded by FM3 to the active tip-tilt mirror, TTM, followed by the low-order (LODM) and high-order (HODM) deformable mirrors, the latter of which resides conjugate to the entrance pupil (the telescope primary is the aperture stop).  A fourth fold, FM4, sends light directly to a matched, re-imaging OAP2.  In converging space, wavefront sensing light is split by an exchangeable dichroic/beam-splitter in reflection at the SSM1 star selection mirror, making a periscope pair with the complementary SSM2, before entering the focusable wavefront sensor, WFS, and acquisition camera assembly, ACam.  Science light proceeds in transmission from SSM1 into the science instrument volume, where each back-end science instrument implements additional optics as necessary.}
\label{p3k_layout} 
\end{figure*}

PALM-3000 uses the same pair of matched PALM-241 off-axis parabolic (OAP) mirrors to retain a compact 1-to-1 magnification optical relay~\citep{Dekany89}.  This economical design was adopted because of the need to reuse the PALM-241 deformable mirror within PALM-3000 (Section \ref{DeformableMirrors}) and to respect the volume constraints within the Cassegrain cage environment of the Hale telescope.  These OAPs were specified with $\lambda$/20 peak-to-valley surface quality over 90\% of their clear aperture and have been recoated without refiguring twice since 1997.  Reuse of the OAPs also retains the same 112 mm diameter internal relay pupil.  The optical layout and a description of the PALM-3000 optical paths is shown in Figure \ref{p3k_layout}.

\subsection{Deformable Mirrors}\label{DeformableMirrors}
PALM-3000 uses two Xinetics, Inc. deformable mirrors sequentially within a collimated optical space to apply appropriate optical phase corrections.  This architectural choice was necessitated by the relatively low 1.06 $\micron$ actuator stroke of the dense high-order deformable mirror relative to the turbulence-induced requirement for total surface stroke ($\pm 5 \sigma$) of 4.1~$\micron$ in conditions where $r_0 = 9.2$~cm~\citep{Hardy98} at $\lambda$ = 500~nm, corresponding to seeing of 1\farcs12 full-width at half-maximum.  The combination of high-spatial-frequency, small-stroke correction from the high-order deformable mirror (HODM) and lower-spatial-frequency, larger-stroke correction from the low-order deformable mirror (LODM) allows PALM-3000 to correct most atmospheric conditions observed at Palomar Mountain (in the worst seeing conditions, the finite total mirror stroke available eventually limits correction capability.)  Summary properties of the high-order deformable mirror and low-order deformable mirror are shown in Table \ref{tab:DMs}.

\subsubsection{Low-order deformable mirror}
The PALM-3000 LODM is the original PALM-241 DM349~\citep{1997SPIE.3126..569O}, which has been in routine use since 1999.  It was stripped and recoated on one occasion in 2003, an operation that involved removal of the mirror facesheet.  In terms of operational reliability, this mirror has seen approximately 500 nights of scheduled science usage.  Assuming 50\% open-shutter time (year-round) and average operation at 1kHz update rate, this mirror has seen $\approx$ 7.5~Gcycles $\times$ 241 active actuators $\approx$ 1.8 trillion actuator-updates in science use.  We estimate that including laboratory time, $\approx$ 8.4 trillion actuator-updates have been made without a single failure at the actuator level\footnote{During 12 years of routine science operation, PALM-241 did on two different occasions suffer broken pins on the integrated 440-pin LODM cable connector, each resulting in the loss of function of one actuator for a period of several weeks.}.  This reliability data, along with similarly experiences at other observatories, is now able to inform design decisions for future generation astronomical AO systems.

\subsubsection{High-order deformable mirror}
The HODM is the largest format astronomical deformable mirror to date, having 66 $\times$ 66 physical actuators organized into 6 $\times$ 6 actuator modules containing 11 $\times$ 11 actuators each.  The actuators are made of electrostrictive lead magnesium niobate (PMN) and are spaced by 1.8~mm center-to-center with 0.15~mm gaps between the actuators.  A 0.2~mm thick glass facesheet is attached to the actuators. The mirror has 3,388 active actuators within a circular clear aperture of 117~mm in diameter.  The AO relay produces a 114-mm pupil which fully illuminates 63 wavefront sensor subapertures when used in $s_{64}$ mode (spanning 64 HODM actuators in a pupil-matching Fried geometry).  The entire DM housing (excluding external wiring) is 23 $\times$ 23 $\times$ 13.2~cm.

The nominal 0-100~V range for actuation is limited by the Xinetics-provided DM drive electronics to $\pm$ 30~V about a +40~V bias signal, which was found experimentally to put the HODM near the center of its linear range.  The drive electronics overall require 55 A of continuous current, 90~A peak, necessitating the installation of auxiliary 120~V power capacity to the Cassegrain environment.

During acceptance testing, the best controlled HODM self-flattening achieved was 11~nm RMS of residual surface error~\citep{2010SPIE.7736E..81R}.  Assuming that spatial frequencies of 8 cycles or less across the pupil are corrected by the LODM, then the HODM stroke required to self-flatten, in the lab, was achieved using less than 5\% RMS of the available HODM stroke, with a few individual HODM actuators using up to 25\% of their range.  

Although the HODM represents a significant prototyping achievement for Xinetics, Inc. there were imperfections created during the manufacturing process which affect the surface figure.  Because the actuators are closely spaced and made of an electrically active material, the entire grid of actuators was dipped in a conformal coating prior to attaching the facesheet.  During initial polish of the completed mirror, polishing slurry leaked into the area between the actuators, causing either the conformal coating or the facesheet bonding adhesive to absorb moisture and wick between the actuators.  When sent to vacuum aluminization, the offensive material dried out and was fixed in place between actuators. Thus, as the material absorbs or releases moisture, it expands and contracts.  Because the actuators are rigidly attached, the inter-actuator dimensional change results in highly local deformations in the facesheet.  The stroke of the majority the actuators has not been compromised, but for a group located along the bottom edge of the HODM, 18 actuators have been found to be restricted to less than half the nominal actuator stroke (8 of which are inscribed within our useful pupil).

A complete solution to this problem would require removal of the facesheet and an attempt to clean the unwanted material from between the actuators.  In addition to risking the facesheet, this would have risked permanent damage the actuators.  As an alternative mitigation, we elected to have the facesheet repolished under strict humidity control and implemented a passive relative humidity (Rh) control system, sealing the HODM case and using a specially-formulated passive, two-way 30\% Rh maintaining Humidipak (now Boveda) coupled to an Rh monitoring system.  Results from initial winter operation indicate the HODM puckers may additionally be temperature dependent, even with controlled Rh, but this remains under investigation.  

During early science observations, the high-spatial-frequency nature of the HODM deformations has been confirmed to move additional stellar light outside the HODM control region, beyond 1\farcs6). Wave-optics simulations on the impact to contrast due to frequency folding effects predict residual stellar speckles to have a contrast ratio of less than $10^{-7}$.

\begin{deluxetable*}{lcc}
\tabletypesize{\scriptsize}
\tablecaption{PALM-3000 Deformable Mirrors}
\tablewidth{0pt}
\tablehead{
\colhead{Parameter} & \colhead{Low-order Deformable Mirror (LODM)} & \colhead{High-order Deformable Mirror (HODM)} }
\startdata
Total Number of Actuators & 349 & 4,356 \\
Number of Active Actuators & 241 & 3,388 \\
Number of Non-Functional Actuators & 0 & 1 \\
Actuator Material & $\begin{array}{c}\text{Electrostrictive Poled} \\ \text{Lead Magnesium Niobate (PMN)} \end{array}$ & Electrostrictive PMN \\
Construction Type & Close-packed Discrete Actuators & $\begin{array}{c}\text{6x6 Mosaic of 11x11-actuator} \\ \text{Integrated Photonex Modules} \end{array}$ \\
P-V Surface Stroke & 4.8 $\micron$ & 1.06 $\micron$ \\
P-V Interactuator Stroke & 1 $\micron$ & 0.5 $\micron$ \\
Available Atmospheric Stroke & $\approx 3 \micron$ & $\approx 1 \micron$ \\
Self-corrected Surface Figure & 19~nm RMS & 11~nm RMS \\
Actuator Linearity over 80\% of Stroke & 15\% & 25\% (bias-dependent) \\
Actuator Pitch & 7.0 mm & 1.78 mm \\
Projected Actuator Pitch On-Sky & 31.25 cm & 8.1 cm  \\
Angle of Incidence of Beam & 16.05 deg & 10.50 deg \\
Facesheet Type & 1.4 mm Zerodur & 0.2 mm Zerodur \\
Facesheet Coating & Protected Aluminum & Protected Aluminum \\
Mirror Assembly Mass & 8 kg & 9 kg \\
Drive Electronics & 0-100 V JPL Custom & 0 - 400 V Xinetics-provided \\
Slaving & 108 Electronically-slaved Perimeter Actuators & None \\
Optical Conjugate Altitude & 780 m & 0 m \\
Delivery Date & 1998 & 2010 \\
Notes & $\begin{array}{c} \text{Up to 15 percent hysteresis at {\raise.17ex\hbox{$\scriptstyle\sim$}} 0C} \\ \text{Overdrive protected via Zener network} \end{array}$ & $\begin{array}{c} \text{Static figure humidity dependent} \\ \text{Overdrive protected via software} \end{array}$
\enddata
%\footnotetext[1]{footnote here.}
\label{tab:DMs}
\end{deluxetable*}

\begin{figure}[h!]
\center
\includegraphics[width=3.0in]{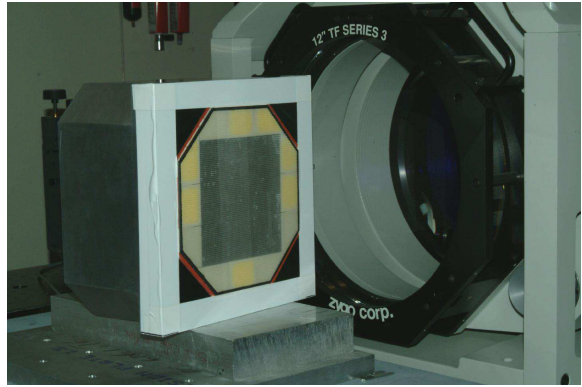}
\caption{ The 66 x 66 actuator HODM, fabricated by Xinetics, Inc. for the PALM-3000 adaptive optics system, undergoing test during optical polishing.}
\label{HODM_pic} 
\end{figure}

\subsection{Wavefront Sensor}\label{WavefrontSensor}
PALM-3000 uses a Shack-Hartmann wavefront sensor with four different pupil sampling formats implemented through a lenslet array exchanger \citep{2008SPIE.7015E.152B}.  The reconfigurable approach, previously adopted for the VLT NAOS system ~\citep{2003SPIE.4839..140R}, allows performance optimization of PALM-3000 for natural guide stars spanning over 18 magnitudes of brightness \citep{Baranec:08}. The Shack-Hartmann detector is a $128 \times 128$ pixel split frame transfer E2V CCD50 packaged into a SciMeasure Analytical Systems ``Li'l Joe'' controller head, a platform originally developed for PALM-241 \citep{2000ASSL..252..395D}.
The camera pixels can be read out at multiple selectable rates which correspond to the full-frame readout rates, with corresponding read-noise values, presented in Table~\ref{tab:HOWFS-CCD}. To enable an effectively continuous selectable frame rate, the detector controller can be programmed with an adjustable delay time between the readout of the last pixel in the frame storage area and the execution of the frame transfer. This allows for slower frame rates than presented in Table~\ref{tab:HOWFS-CCD} while still operating at one at the four pixel rates and read-noise values.

\subsubsection{Pupil Sampling Modes}
We index the four PALM-3000 pupil sampling modes according to the one-dimensional number of Shack-Hartmann subapertures formed across the pupil diameter, $s$~=~\{64,~32,~16,~8\}\footnote{For simplicity, we hereafter will refer to the physically 63-across pupil sampling mode as s64 mode.}.  Each subaperture mode operates in a different diffraction regime and thus creates different subaperture point spread functions (PSFs).  The $s_{64}$ PSFs are dominated by diffraction with large 2\farcs1/pixel quad-cells, making centroiding precision particularly sensitive to CCD charge diffusion.  The $s_{64}$ plate scale was chosen to give a ratio of pixel to spot size of $\approx$ 1, providing reasonable linearity around the quad-cell zero-point \citep{Hardy98}. The plate scale for $s_{32}$ was similarly chosen for linearity if using only $2 \times 2$ pixels in NGS  mode; however, when guiding on extended objects $s_{32}$ is intended to calculate a center of mass using $4 \times 4$ pixels, though full implementation of this is not yet complete.  Were the pixel-to-image size ratio for $s_8$ mode maintained at $\approx$ 1, the $f$/\# of the microlens array would be reduced to the point where aberrations in the AO relay would be the dominant factor in spot size, increasing centroiding error unnecessarily. To compensate for this, we adopted the same $f$/\# for $s_8$ as we selected for $s_{16}$, with the result that the $s_8$ plate scale is finer than that for $s_{16}$ by a factor of two.

\begin{deluxetable*}{lccccc}
\tabletypesize{\tiny}
\tablecaption{PALM-3000 Wavefront Sensor}
\tablewidth{0pt}
\tablehead{
\colhead{Parameter} & \multicolumn{5}{c}{Wavefront Sensor Mode} \\
& $s_{64}$ & $s_{32}$ & $s_{16}$ & $s_8$ & $s_1$}
%\colhead{Parameter} & \colhead{S64} & \colhead{S32} & \colhead{S16} & \colhead{S8} }
\startdata
Subapertures per Pupil Dia. & 63 & 32 & 16 & 8 & 1 \\
No. of Majority-Illuminated Subapertures & 2804 & 708 & 177 & 48 & 1 \\
Projected Subaperture Diameter [cm] & 8.06 & 15.88 & 31.75 & 63.5 & 508 \\
\hline
Pixel Scale On-Sky ["/pixel] & 2.1 & 1.5 & 1.3 & 0.65 & -- \\
Subimage Spacing [pixels] & 2 & 4 & 8 & 16 & -- \\
Subaperture Field of View ["] & \multicolumn{4}{c}{0.48 -- 3.84, variable\tablenotemark{1}} & -- \\
%Subaperture Field of View ["]\tablenotemark{1} & 4.2 & 6.0 & 10.4 & 10.4 & -- \\
Design Subaperture Polychromatic Image FWHM ["] & 2.1 & 1.5 & 1.3 & 1.3 & --\\ 
\hline
Lenslet Substrate & \multicolumn{4}{c}{$\text{SiO}_2$} & -- \\
Lenslet Pitch [$\micron$] & 150 & 300 & 600 & 1200 & --\\
Lenslet Focal Length [mm] & 14.05 & 19.66 & 22.69 & 45.38 & $\infty$ \\
Lenslet Radius of Curvature [mm] & 6.32 $\pm 2.3\%$ & 8.85 & 10.21 & 20.42 & -- \\
Lenslet Surface Sag [$\micron$] & 0.5 & 1.3 & 4.5 & 8.9 & --\\
Lenslet Focal Ratio & 93.0 & 65.1 & 37.6 & 37.6 & --\\
\hline
Nominal Spatial Filter Setting ["] & 1.92 & 0.96 & 0.48 & -- & 3.84\\
Nominal Spatial Filter Setting [$\micron$] & 760 & 380 & 190 & -- & 1560\\
\hline
Subaperture Pixel Modes \tablenotemark{2} & 2x2 & $\begin{array}{c} \text{c 2x2} \\ \text{4x4 b 2x2 \tablenotemark{3}} \\ \text{4x4} \end{array}$ & $\begin{array}{c} \text{c 2x2} \\ \text{c 4x4 b 2x2  \tablenotemark{3}} \\ \text{c 4x4} \end{array}$ & $\begin{array}{c} \text{c 4x4} \\ \text{c 4x4 b 2x2  \tablenotemark{3}} \\ \text{c 8x8} \\ \text{c 8x8 b 2x2  \tablenotemark{3}} \end{array}$ & 128x128 \\
\hline
WFS Geometry & Fried & Non-Fried & Non-Fried & Non-Fried & -- \\
\hline
Notes & -- & -- & $\begin{array}{c} \text{Low priority} \end{array}$ & -- & $\begin{array}{c} \text{Pupil Img Only} \end{array}$
\enddata
\tablenotetext{1}{Set by a variable-width square field stop / spatial filter (\S\ref{sec:spatial_filter}).}
\tablenotetext{2}{c = \textquotedblleft central" (e.g. c 2x2 = central 2x2 physical pixels from larger subaperture); b = \textquotedblleft binned down to" (e.g. 4x4 b 2x2 = 4x4 physical pixels binned on-chip down to 2x2 pixels at readout).}
\tablenotetext{3}{As of this writing, binning modes have yet been implemented due to emphasis to optimize high-contrast observations on bright stellar targets.}
\label{tab:HOWFS}
\end{deluxetable*}

Foreshortening of the pupil as projected onto the HODM due to the 10.5-degree angle of incidence by the chief ray is compensated by the wavefront sensor optics to provide true Fried geometry \citep{Fried:77}.  The optical conjugate of the LODM is at a point approximately 780 meters above the observatory and although the system is operated in pseudo-single-conjugate mode, some astrometric degradation over the full 40\arcsec $\times$ 40\arcsec ~PHARO field is tolerated.  Although non-pupil deformable mirror imperfect motions increase the stochastic error, and thus require additional averaging to overcome noise sources, the systematic errors that arise from variations in slowly-varying component of LODM shape are the most problematic for precision astrometry.  In an experiment we measured the separation of a binary star pair with both a flattened LODM and after applying the full LODM stroke.  Over the approximately 20 arc sec binary separation, we determined the LODM shape induced a 100~milliarcsec shift, about 0.5\%.  Over the much smaller exoplanet work angle, and for reasonably long-time-average LODM figure drift of 150~nm P-V, we estimate the systematic astrometric errors from our choice of LODM conjugation should be less than 600~microarcseconds ($\mu$as).  In the long run, we expect extremely precise binary star orbits derived from GAIA data to allow us to calibrate the long-term systematic astrometric errors induced by our two-deformable-mirror architecture to better than 40~$\mu$.

\subsubsection{Spatial Filter}
\label{sec:spatial_filter}
The PALM-3000 wavefront sensor includes a variable aperture optical spatial filter \citep{2004JOSAA..21..810P, Swartzlander08} that serves as the field stop, limiting crosstalk between sub-aperture detector pixels. It is also designed to mitigate wavefront aliasing errors that result in lower AO performance which can corrupt high-contrast observations.  The spatial filter used in PALM-3000 has a square aperture ranging from 190 to 1560 $\mu$m, with the largest aperture corresponding to 3\farcs84 which allows the extended objects Uranus and Neptune to be used as AO guide sources.  The appropriate aperture sizes for high-contrast use are shown in Table~\ref{tab:HOWFS}.  We do not intend to operate the spatial filter for anti-aliasing in $s_{8}$ mode, where the contribution to either contrast or wavefront errors is generally small.

\begin{deluxetable*}{lcc}
\tabletypesize{\scriptsize}
\tablecaption{PALM-3000 Wavefront Sensor Detector}
\tablewidth{0pt}
\tablehead{\colhead{Parameter} & & \colhead{Value} }
\startdata
Detector Type & & E2V CCD50 \\
Pixel Size & & 24 $\micron$ \\
Camera Manufacturer & & SciMeasure Analytical Systems, Inc. \\
Dark Current (278 K ambient) & & 150 - 780 e- / pixel / sec \\
\hline
$\begin{array}{c} \text{Readout rate} \\ 2000 \text{ fps} \\ 1200 \text{ fps} \\ 500 \text{ fps} \\ 200 \text{ fps} \end{array}$ & 
$\begin{array}{c} \text{Readout noise} \\ 9.3 e- \\ 7.3 e- \\ 3.9 e- \\ 3.1 e- \end{array}$ & 
$\begin{array}{c} \text{Camera gain} \\ 0.1 e- / DN \\ 0.4 e- / DN \\ 0.7 e- / DN \\ 2.5 e- / DN \end{array}$ \\
\hline
\enddata
\label{tab:HOWFS-CCD}
\end{deluxetable*}

\subsection{Mechanics and Interfaces}
To minimize the impact on observatory operations and avoid unnecessary engineering, we reused the main optical bench, its handling cart and rotisserie fixture, and internal telescope and source simulator from PALM-241.  Thus, the PALM-3000 installs horizontally at Cassegrain focus, with the AO relay optics suspended from the underside of the optical bench.  Earlier finite element analyses, verified by PALM-241 flexure tests on the turnover fixture, demonstrated a total 180-degree inversion deflection of the bench surface of approximately 1mm (in this configuration, the bench is held by its endpoints within the rotisserie fixture).  No appreciably local flexure of the bench face sheet due to mounted optics has been seen to date, although we take care to distribute the load from science instruments over a total face sheet area of $\approx$ 230 cm$^2$.  We currently limit science instruments to a volume spanning the 137 cm width of the AO bench $\times$ 45.7 cm depth (measured along a radius extending from the Cassegrain rotation axis) $\times$ 91.4 cm height perpendicular (and hanging below) the lower bench surface.  A mass limit of 200 kg is set by face sheet safety margin while an axial torque limit of 34.6 m-kg about the telescope axis is imposed to avoid back-driving of the Cassegrain ring rotation motor.  The total mass installed onto the Cassegrain cage during a PALM-3000 setup with the heaviest instrument (P1640) is $\approx$ 3,230 kg, which has required additional counterweights to be added to the top of the telescope ring for balance.  Despite the large weight, no detriment to any telescope performance parameter has been noted.

\subsection{Real-time Computation }
\subsubsection{Reconstructor pipeline and wavefront control}

The large number of DM actuators in the PALM-3000 system requires an accurate reconstructor matrix, mapping centroid measurements to wavefront phase, in order to achieve robust closed-loop stability ~\citep{2001JSV...246..281B}.  The system experiences significant pupil shifts in response to changes in telescope declination, thus the reconstructor must be periodically re-identified throughout the course of an observing period. To facilitate this process, PALM-3000 possesses a unified reconstructor pipeline that autonomously gathers current telescope data and computes a new reconstructor matrix. 

The telescope operator initiates the computation by calling the reconstructor pipeline script. The script is structured such that pertinent reconstructor parameters can be specified if desired, however the default parameters are typically sufficient and no further input from the user is required.  The pipeline begins by gathering and processing basic observing information, such as telescope position and subaperture flux measurements. The reconstructor matrix itself is the result of a regularized and weighted pseudo-inverse performed on precomputed actuator influence matrices. The weighting and regularization terms incorporate pupil illumination information, as well as {\em a~priori} Kolmogorov phase covariance values that may be adjusted based on the prevailing seeing~\citep{Law:1996}. The inversion calculation is computationally intensive, thus the HODM and LODM reconstructor computations may be performed together or separately on remote servers or CPU cores. Once the HODM and LODM reconstructors are computed, they are combined into a single matrix and automatically loaded directly into the PALM-3000 servo controller. Two additional rows are added to account for tip/tilt residuals. An optional ``offload" matrix is also computed, which projects HODM actuator positions to the space of LODM actuators.

During each control loop iteration, the PALM-3000 real-time computer multiplies centroid measurements with the reconstructor matrix to provide a residual phase estimate in DM actuator space. The wavefront controller consists of parallel ``leaky" integrators for the HODM, LODM and fast-steering mirror, which act on the residuals to generate position commands. The fast-steering mirror alone does not provide the necessary bandwidth to fully compensate for tilt/tip errors, thus a portion of these residuals are offloaded to the LODM. To harness the large stroke range of the LODM, an alternative control scheme offloads HODM residuals to the LODM integrator, effectively creating a double-intergral controller for low order aberrations. Control gains can be adjusted during closed-loop operation to optimize system performance and maintain stability.

\subsubsection{Real-time computer}
The PALM-3000 real-time computer (RTC) for wavefront reconstruction is based on a GPU implementation of 16 retail NVIDIA 8800 GTX graphics cards distributed over eight dual-core Opteron PCs from Hewlett-Packard, each hosting two cards~\citep{2008SPIE.7015E..95T} in conjunction with a Servo PC that combines partial reconstruction results and a Database PC which provides high-speed telemetry.  The GPU architecture provides supercomputing-like power and memory bandwidth coupled with ease of programming through a low-level C interface. All 10 PCs are interconnected using a Quadrics QsNetII 16-port switch that delivers over 900 MB/s of user space to user space bandwidth each direction with latency under 2 microseconds for a total of 14.4 GB/s of bisectional bandwidth and broadcast capability.  High-bandwidth AO telemetry data, such as wavefront sensor pixel data, as well as latency-sensitive data such as HODM commands, are transferred using fiber optics.  Low-bandwidth data, such as acquisition camera pixels and electronics hardware status, are sent from the Servo PC to the Database PC via a dedicated 1Gbit Ethernet.  Long latency-tolerant commands, such as those issued by the Servo PC to stepper motors and lamps are sent via direct connections. 

The measured mean processing latency of the wavefront reconstruction based on a full matrix vector multiplication using the input matrix of 4096 actuator values by 8192 centroid values and 2 $\times$ 2 pixel centroiding has been measured to consist of the following terms: 94 $\mu$s for centroiding and vector-matrix-multiply in parallel on 8 PCs, 41 $\mu$s transfer of partial residuals to combining PC, 70 $\mu$s for servo calculation setting the new actuator commands, 35 $\mu$s for software inter-actuator voltage check to protect the HODM from potentially damaging strain (reducible to 10 $\mu$s via faster algorithm), 125 $\mu$s communication through the DM interface board (reducible to 25 $\mu$s by reimplementing the Curtiss-Wright driver ourselves), and 50 $\mu$s inter-actuator voltage check in FPGA hardware.  Start-of-data to end-of-data delay was confirmed at 475 $\mu$s +- 25 $\mu$s jitter.  In addition 30 $\mu$s of frame transfer delay in the wavefront sensor controller. Latency is measured from the falling edge of Start of Frame signal (e.g. the end of frame transition) on the Little Joe Camera transmitting CameraLink card, to the 50\% transition point of an actuator's drive voltage and does not include the camera integration time.  Measurement via oscilloscope has shown this latency to be 956 microseconds for 1,250 - 2,000 Hz frame rate readout and 1280 microseconds for camera readout in the range of 300 - 1249 Hz.  The mean compute latency in the real-time controller has been benchmarked at 415 microseconds, with standard deviation of 30 microseconds.

The HODM driver is specified for a 3 A current limit.  The rise time of a single actuator, settling to within 5\% of a 300~nm step function, was measured to be 20 $\mu$s, consistent with this current limit.  The HODM driver latency, measured from the sending of a digital command to the setting of the actuator voltage, was found to be 60 $\mu$s.  Both of these measurements confirm the design specifications of our drive electronics.

% The instrument control software architecture is built upon many structuring concepts and execution model of the TinyOS operating system, namely, a component-based architecture, structured event-driven execution model, and split-phase operations.  Like TinyOS, the PALM-3000 software system uses active messages for the communication method and is written entirely in the component-oriented language of nesC.  It achieves its distributed processing through the use of bidirectional proxy components that are interposed between user components and remote provider components.

\subsubsection{Wavefront Calibration}
Non-common-path static wavefront  calibration for PALM-3000 is implemented via an interface allowing for the loading of element slope offset vectors in the appropriate pupil mode which represents the target command point in the real-time AO wavefront reconstruction.  These slope offsets, represented internally in Shack-Hartmann sensor pixel coordinates, are generated via different techniques depending on the science camera undergoing calibration. For PHARO, we use a modified iterative Gerscherg-Saxton phase diversity algorithm to achieve wavefront calibration to better than 40~nm RMS phase error \citep{2012SPIE.8447E..0YR, 2010SPIE.7736E.197B}.  For P1640, an electric field measuring calibration interferometer has so far achieved approximately 14~nm wavefront error \citep{Vasisht:13}.

\subsection{Cooling Infrastructure}
Prior to PALM-3000 development, the Hale Telescope at Palomar had no capacity to provide instrumentation with facility liquid cooling.  Experimental investigation of the PALM-241 system, which dissipated a total of 1.5 kW below the primary mirror determined this heat flow led to an approximately $1^{\circ}$C temperature rise in the area of the primary mirror directly above the dissipating electronics.  While not directly detrimental to PALM-241 observations, the thermal inertia of the primary mirror demonstrated residual surface distortions for up to 40 hours, degrading some non-AO operations, particularly after comprehensive refurbishment of the Hale primary mirror back supports improved seeing-limited image quality in 2009.

As the PALM-3000 AO system dissipates approximately 7 kW during HODM operation, the project and Palomar staff jointly undertook the addition of a liquid cooling system to remove heat from the electronics in the Cassegrain cage area.  The new cooling system comprises a primary facility chiller in conjunction with a secondary process chiller, which circulates coolant to multiple fan tray heat exchangers operating within each rack.  The facility chiller ejects its heat through an exhaust tunnel which is vented a sufficient distance from the telescope dome.  The diluted glycol coolant is currently conveyed to the Cassegrain cage by way of a draped umbilical cord, which also provides additional electrical power and fiber data lines required for PALM-3000.  Despite the additional torque of the draped umbilical, no noticeable change to the telescope pointing performance has been found after a routine pointing model recalibration.
    
\section{Instrument Suite}
PALM-3000 has thus far been commissioned with five science instruments as summarized in Table~\ref{tab:Instruments}.  The range of science capability provided by the diverse PALM-3000 instrument suite includes direct near-infrared and visible imaging, slit-based near-infrared grism spectroscopy, moderate visible resolution and low near-infrared resolution integral field spectroscopy, coronagraphy, and near-infrared nulling interferometry.  Fast-frame region-of-interest array imaging is also available in both the visible and near-infrared.

\begin{deluxetable*}{lcccccc}
\tabletypesize{\scriptsize}
\setlength{\tabcolsep}{0.01in}
\tablecaption{PALM-3000 commissioned instrument suite}
\tablewidth{0pt}
\tablehead{
\colhead{Instrument} & \colhead{Builder} & \colhead{$\begin{array}{c}\text{First} \\ \text{Light} \end{array}$} & \colhead{Type} & \colhead{$\begin{array}{c}\text{Spectral} \\ \text{Range} \\ \text{[$\micron$]} \\ \end{array}$} & \colhead{$\begin{array}{c}\text{Spectral} \\ \text{Resolution}, \frac{\delta \lambda}{\lambda} \end{array}$} & \colhead{$\begin{array}{c}\text{Spatial} \\ \text{Sampling} \\ \text{[mas/pixel]} \end{array}$} }
\startdata
\multirow{5}{*}{PHARO \tablenotemark{a}} & \multirow{3}{*}{Cornell U.} & \multirow{3}{*}{1999} & 1k x 1k HgCdTe Imager & \multirow{5}{*}{0.97 - 2.4} & 5 - 100 & \multirow{5}{*} {25, 40} \\
& & & Grism Spectrograph & & 1,310 - 2,110 & \\
& & & Lyot Coronagraph & & 5 - 100 & \\
& Caltech & 2009 & Band-limited-mask Coronagraph & & {\raise.17ex\hbox{$\scriptstyle\sim$}} 5 & \\
& JPL & 2009 & Vector-vortex Coronagraph & & {\raise.17ex\hbox{$\scriptstyle\sim$}} 5 & \\
\hline
PFN \tablenotemark{b} & JPL & 2007 & $\begin{array}{c}\text{Fiber Nullling} \\ \text{Interferometer} \end{array}$ & 2.2 & {\raise.17ex\hbox{$\scriptstyle\sim$}} 5 & IWA = 30 mas \\
\hline
SWIFT \tablenotemark{c} & Oxford U. & 2009 & $\begin{array}{c}\text{44 x 89 spaxel Integral-} \\ \text{Field Spectrograph (IFS)} \end{array}$ & 0.60 - 0.90 & {\raise.17ex\hbox{$\scriptstyle\sim$}} 3,250 - 4,400 & 80, 160, 235\\
\hline
P1640 \tablenotemark{d} & $\begin{array}{c}\text{AMNH} \\ \text{JPL} \end{array}$ & 2009 & ${\begin{array}{c} \text{200 x 200 IFS Coronagraph} \\ \text{with nm-Level Metrology} \end{array}}$ & 1.06 - 1.78 & 33 - 58 & 19.2 \\
\hline
\multirow{3}{*}{TMAS \tablenotemark{e}} & \multirow{3}{*}{Caltech} & \multirow{3}{*}{2012} & 2560 x 2160 sCMOS Imager & \multirow{2}{*}{0.38 - 0.90} & \multirow{2}{*}{5 - 100} & \multirow{3}{*} {10, 16} \\
& & & Vector-vortex Coronagraph (planned) & & & \\
& & & Eyepiece (planned) & {0.39 - 0.70} & millions\tablenotemark{f} &
\enddata
\tablenotetext{a}{\cite{2001PASP..113..105H}}
\tablenotetext{b}{\cite{2011ApJ...729..110H}}
\tablenotetext{c}{\cite{2010SPIE.7735E.258T}}
\tablenotetext{d}{\cite{2011PASP..123...74H}}
\tablenotetext{e}{R. Dekany et al. (2013, in preparation)}
\tablenotetext{f}{\cite{Geldard72}}
\label{tab:Instruments}
\end{deluxetable*}

\section{Predicted Performance}
\subsection{Error Budgets}
\label{sec:predicted_performance}
PALM-3000 is capable of minimizing the residual wavefront error for guide stars that range in brightness by over a factor of $10^7$ through adjusting the combination of wavefront sensor pupil sampling, camera frame rate, and servo control modes.  Naturally, the delivered level of wavefront correction over this broad range of guide star brightness varies significantly.  Insight into the intrinsic wavefront error arising from input guide star photon noise, temporal delay due to finite integration time, and finite spatial correction bandwidth as a function of theoretical continuous pupil sampling density, $s$ (samples per pupil diameter), is shown in Figure \ref{tab:Performanceprediction}.

PALM-3000 has adopted a strategy of performing the highest bandwidth tip-tilt correction using the relatively limited stroke of the LODM, coupled to a moderate-speed offload through the servo control law to the independent, large-stroke tip-tilt mirror.  The  -3db tip-tilt rejection bandwidth of the tip-tilt mirror direct loop has been measured to be 18 Hz, while that of the cascaded approach is expected upon full implementation to achieve $>$ 40 Hz.  Because the typical fast stroke usage of the LODM for tip-tilt correction is only a few hundred nanometers RMS, there is generally little impact to the available stroke for higher-mode corrections, though the impact of spurious tilt events to LODM stability remains under investigation.

\begin{deluxetable*}{lcc}
\tabletypesize{\scriptsize}
\tablecaption{PALM-3000 Predicted Performance Error Budget}
\tablewidth{0pt}
\tablehead{\colhead{RMS Error Term} & \colhead{$\begin{array}{c}\text{High-Contrast Exoplanet Imaging} \\ V = 7 \\ \text{s64 mode} \end{array}$} &  \colhead{$\begin{array}{c}\text{Kepler Candidate Follow-Up} \\ V = 16 \\ \text{s8 mode} \end{array}$}}
\startdata
Atmospheric Fitting Error \tablenotemark{a} [nm] & 44 & 213 \\
Bandwidth Error \tablenotemark{b} [nm]& 46  & 220 \\
Measurement Error \tablenotemark{c} [nm] & 41 & 202 \\
Static Calibration Error [nm] & 30 & 30 \\
Other High-order Errors \tablenotemark{d},\tablenotemark{e} [nm] & 65 & 103 \\
\hline
Total High-order Wavefront Error [nm] & 105 & 382 \\
\hline
Tip-tilt Errors\tablenotemark{f} [mas] & 2.8 & 22 \\
\hline
Predicted Strehl Ratio $\begin{array}{c}\text{V} \\ \text{I} \\ \text{H} \\ \text{K} \end{array}$ & $\begin{array}{c}\text{0.32} \\ \text{0.58} \\ \text{0.88} \\ \text{0.93} \end{array}$ & $\begin{array}{c}\text{--} \\ \text{--} \\ \text{0.08} \\ \text{0.23} \end{array}$
\enddata
\tablenotetext{a}{Atmospheric $r_0 (500 \text{nm})$ = 9.2 cm at $\zeta$ = 10 degrees zenith angle; $\alpha_{DM}$ = .28}
\tablenotetext{b}{Atmospheric $\tau_0 (500 \text{nm})$ = 3.0 ms; bright star 2 kHz frame rate, -3db bandwidth = 100 Hz; faint star 182 Hz frame rate, -3db bandwidth = 12 Hz}
\tablenotetext{c}{Total WFS photodetection efficiency = 0.24}
\tablenotetext{d}{Includes uncorrectable instrument errors, multispectral error, scintillation error, wavefront sensor scintillation error, DM finite stroke and digitization errors, and imperfect anti-aliasing residual error~\citep{Hardy98}; does not include angular anisoplanatism}
\tablenotetext{e}{Includes tip-tilt measurement, bandwidth, centroid anisoplanatism, chromatic dispersion, vibration, and non-common-path mechanical drift errors; no angular anisokineticism}
\tablenotetext{f}{Bright star -3db tip-tilt bandwidth = 35 Hz; faint star -3db tip-tilt bandwidth = 10 Hz}
\label{tab:Performanceprediction}
\end{deluxetable*}

\subsection{Performance vs. Guide Star Magnitude}

\begin{figure*}[t]
 \center
 \includegraphics[width=5.5in]{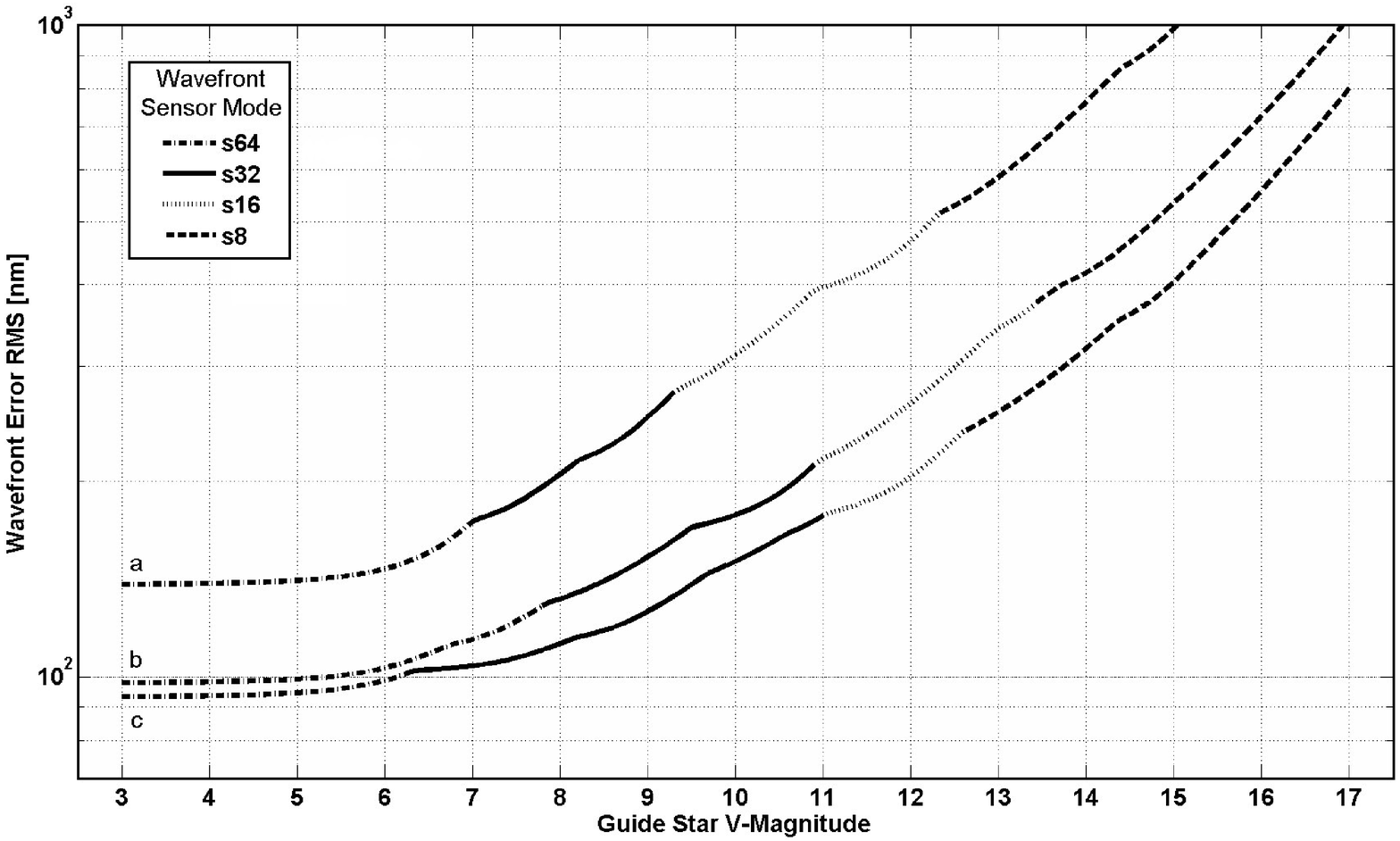}
 \caption{Performance prediction for PALM-3000, with appropriate choice of optimal wavefront sensor mode and high-order wavefront sensor frame rate, under assumed conditions , a)~$r_0$~=~6.0 cm, $r_{0_\text{eff}}$~=~5.5 cm at $t_{0_\text{eff}}$~=~0.79 ms b)~$r_0$~=~9.2 cm, $r_{0_\text{eff}}$~=~8.4 cm, $t_{0_\text{eff}}$~=~2.46 ms, and c)~$r_0$ = 15 cm, $r_{0_\text{eff}}$~=~13.8 cm, $t_{0_\text{eff}}$~=~2.46 ms, where the atmospheric parameters at zenith have been scaled to effective parameters corresponding to 30 degree zenith angle.}
 \label{perf_curve} 
\end{figure*}

As with all natural guide star AO systems, PALM-3000 correction performance is a function of the guide star brightness.  Unlike many other systems, however, the variable Shack-Hartmann pupil sampling of PALM-3000, combined with the array of selectable wavefront sensor camera frame rates, allows for quasi-continuous optimization of performance.  The predicted performance for the optimal choice of pupil and camera setting is shown in Figure \ref{perf_curve}.

\section{Initial Results}
On the nights of June 20-22, 2011, the combination of LODM and HODM correction was initially applied to a stellar source.  In visible ($\lambda = 0.55$ $\micron$) seeing of 0\farcs9, PALM-3000 achieved a first-light K-band Strehl Ratio of 79\% as measured by 30-s exposures in the PHARO imaging camera, exceeding the highest Strehl ever obtained with PALM-241.  This corresponded to an effective wavefront error of 170~nm RMS.  Several recognized limitations were known at the time, such as our having only partially implemented our reconstructor pipeline and not yet fully implemented the MGS calibration algorithm with PHARO.  During the initial 12 months following our PALM-3000 first lock, the AO team was engaged in commissioning back-end science instruments at an average rate of one every three months, limiting the engineering effort available for AO performance optimization.

Since June 2012, PALM-3000 performance has steadily improved through detailed analyses referencing against the design error budget.  This process revealed an unexpected chromatic aberration, lateral color, in the high-order wavefront sensor optical relay.  This chromatic magnification error, which varies with guide star effective temperature, induces an effective achromatic focus error in the science beam equivalent to up to 80~nm RMS wavefront error if uncompensated.  Operationally, we have worked around this issue by using a calibrated focus offset until an optical solution to the lateral color is implemented.  Similarly, small flexure-based pupil illumination changes were discovered to impact performance as a function of zenith pointing.  The prompted our team to improve our reconstructor-generating software to automatically incorporates a measured pupil illumination function for different values of zenith pointing.

\subsection{Current Performance of $s_{64}$ Pupil Sampling Mode}
\begin{figure*}[b]
 \center
\includegraphics[width=5.5in]{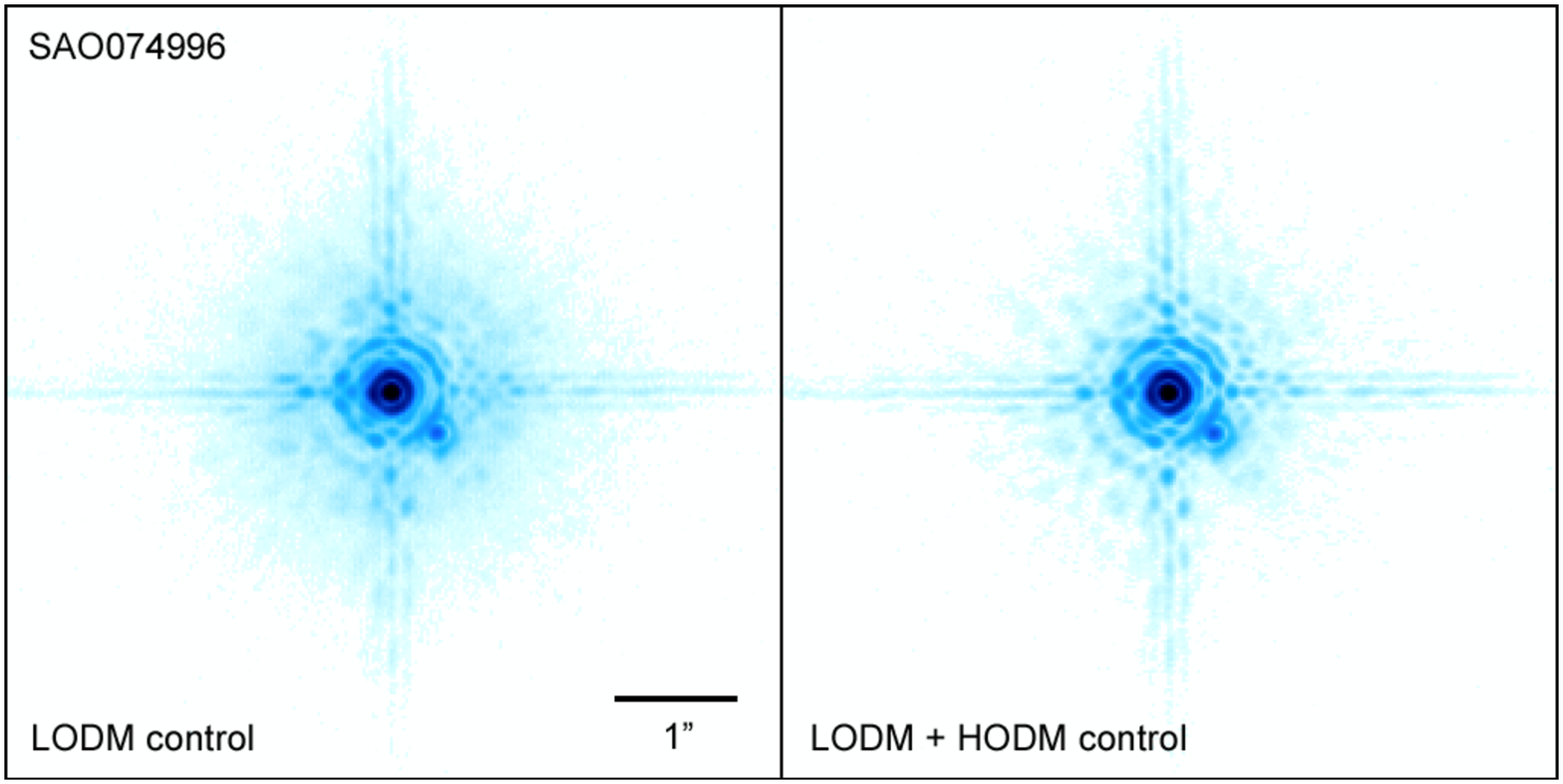}
 \caption{Logarithmic stretch infrared long-exposure observations of SAO074996 controlling only the LODM (left) and the combination of the LODM and HODM (right).  Under LODM control, the working angle for clearing out atmospheric speckles is merely 4 $\lambda$/D.  With additional HODM control, the working angle opens to 32 $\lambda$/D, significantly clearing away the atmospheric halo that underlies the diffraction-induced point spread function.  Note, a ghost reflection, from filters in the PHARO imager, appears to the lower right of the star in these images.}
 \label{LODM_only_PSF} 
\end{figure*}

On-going system optimization has thus far achieved residual wavefront error as low as 141~nm RMS on bright guide stars.  This was achieved on UT 26 September 2012 in s64 mode when guiding on the V = 3.41 star, SAO 074996, and observing near zenith in approximately 1.0 arc sec FWHM seeing.  Figure \ref{LODM_only_PSF} shows a long-exposure image (logarithmic stretch) of this target in the K$_s$ filter, under two operating configurations:  LODM correction only and combined LODM+HODM correction.  Each image consists of a series of PHARO images forming total integration of 272 seconds.  For the LODM-correction-only image, the HODM was set to a fixed optimal figure previously determined using internal calibration sources.  The K$_s$ Strehl ratio is measured using our aforementioned technique to be 72\%.  This is somewhat underperforming our error budget prediction for LODM-only operation.  The second image was formed running PALM-3000 with concurrent LODM and HODM correction, delivering K$_s$ Strehl ratio of 85\%.  The equivalent wavefront error of 141~nm RMS still falls short of the design requirement of 105~nm RMS on bright stars.  We continue to optimize our woofer/tweeter control law and verify parameters following our error budgets.  

In terms of clearing out the point spread function within the IWA, the striking advantage of the HODM control is obvious.  Any coronagraph optimized to reduce image diffraction, particularly in the range between the outer working angle of the LODM (4 $\lambda$/D) and that of the HODM (32 $\lambda$/D), will enjoy the strong benefit of PALM-3000 clearing away residual energy from the seeing halo.  Note, conditions were exceptionally stable on the night of this test and we have no reason to believe conditions varied appreciably in the ~5 minutes that elapsed between these two measurements.

Initial investigations of PALM-3000 performance vs. guide star brightness have been limited to the initially commissioned $s_{64}$ mode, as shown in Figure \ref{perfVbrightness} below.  Between our initial tests in August 2011 and recent performance in September 2012, a number of the initial limitations to performance have been address, although performance has not yet fully reached the error budget expectation and system optimization continues.

\begin{figure*}[h]
 \center
\includegraphics[width=5.5in]{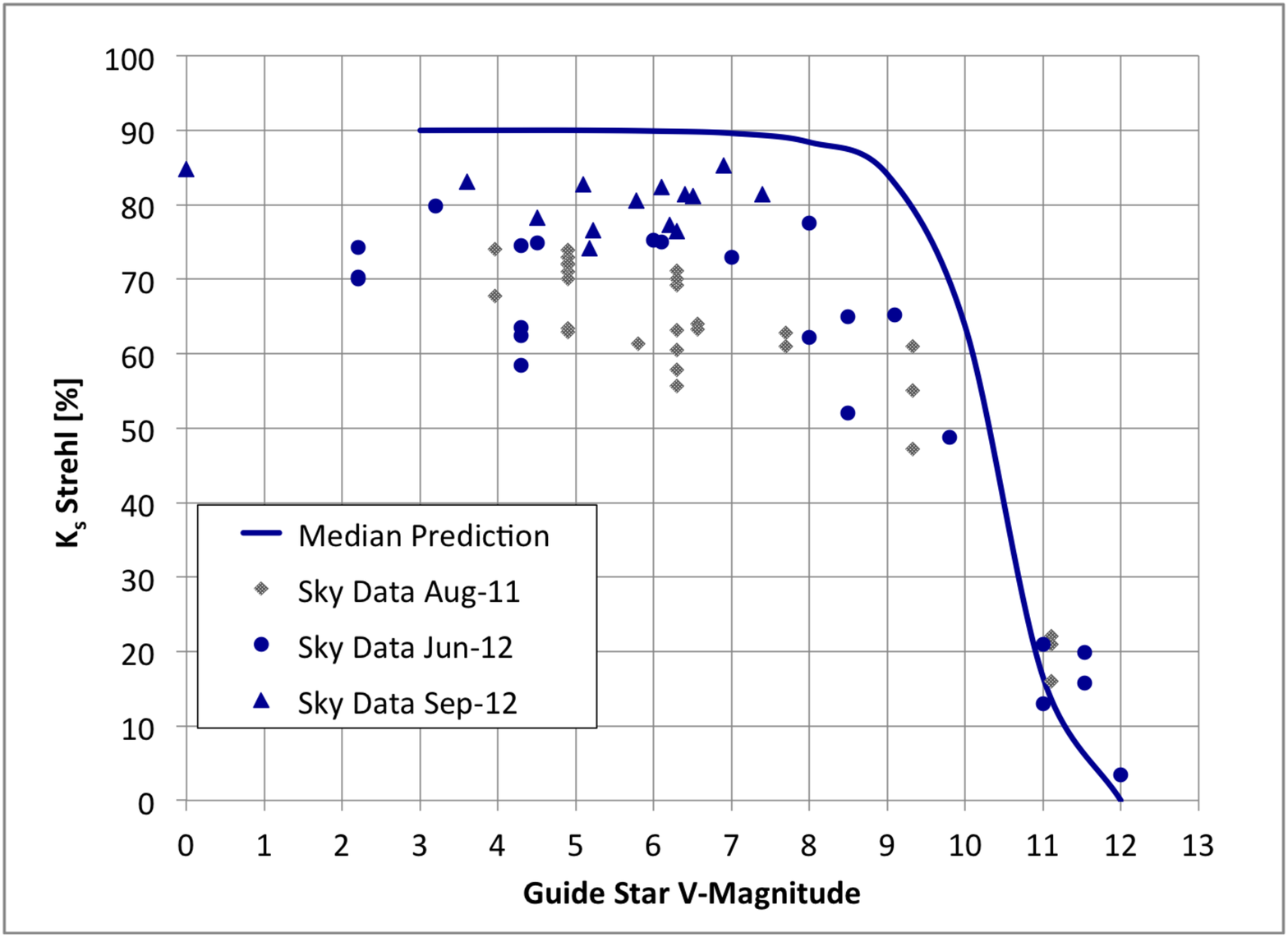}
 \caption{Summary of PALM-3000 $s_{64}$ mode on-sky performance to date as a function of guide star magnitude.  Shown are Strehl performance results in the $K_s$ band as measured by PHARO for a range of guide star brightnesses at two epochs under seeing conditions comparable to median conditions.  For comparison, our error budget prediction for performance is shown as the solid curve.  Note, $s_{64}$ mode is not the optimum setting for guides stars of brightness fainter than V $\approx$ 7 (\S\ref{sec:predicted_performance}), but engineering demonstrations to V $\approx$ 13 to date have validated the faint star performance of the wavefront sensor camera.}
 \label{perfVbrightness} 
\end{figure*}
 
\subsection{Planetary science capability}
The diffraction limited resolution of the 5.1-m Hale Telescope is 16~milliarcseconds at 0.4~$\mu$m wavelength, comparable to the planned Thirty-Meter Telescope resolution in K-band \citep{2003SPIE.4839.1165D} .  Although we expect PALM-3000 to achieve low blue-wavelength Strehl ratio performance in all but the most favorable seeing conditions, significant improvement can be obtained from short-exposure images using post-processing techniques developed at Palomar \citep{2009ApJ...692..924L} and elsewhere \citep{2010A&A...518A...6A, 2013MNRAS.429.1367S}.  Observations of solar system objects at these high angular resolutions, particularly with SWIFT \citep{doi:10.1117/12.925328, 2010SPIE.7735E.258T}, are already underway.

\begin{figure*}[t!]
\center
\includegraphics[width=2.543in]{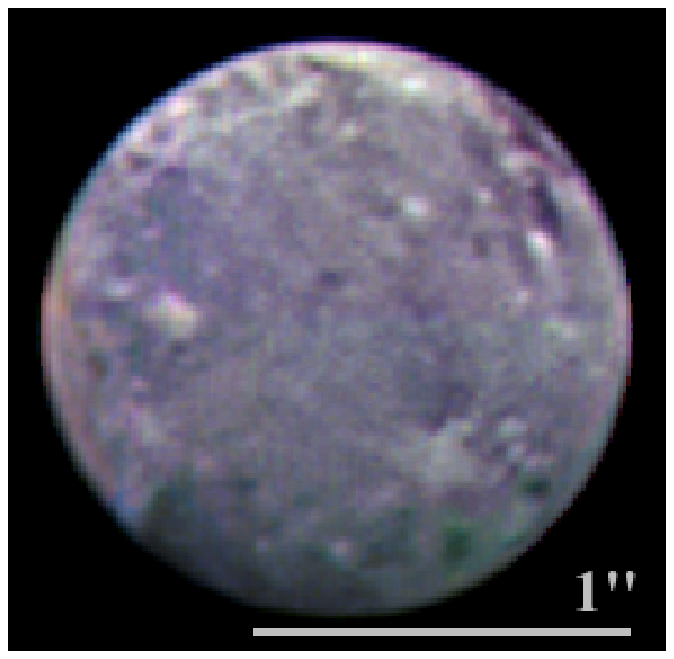}
\includegraphics[width=2.5in]{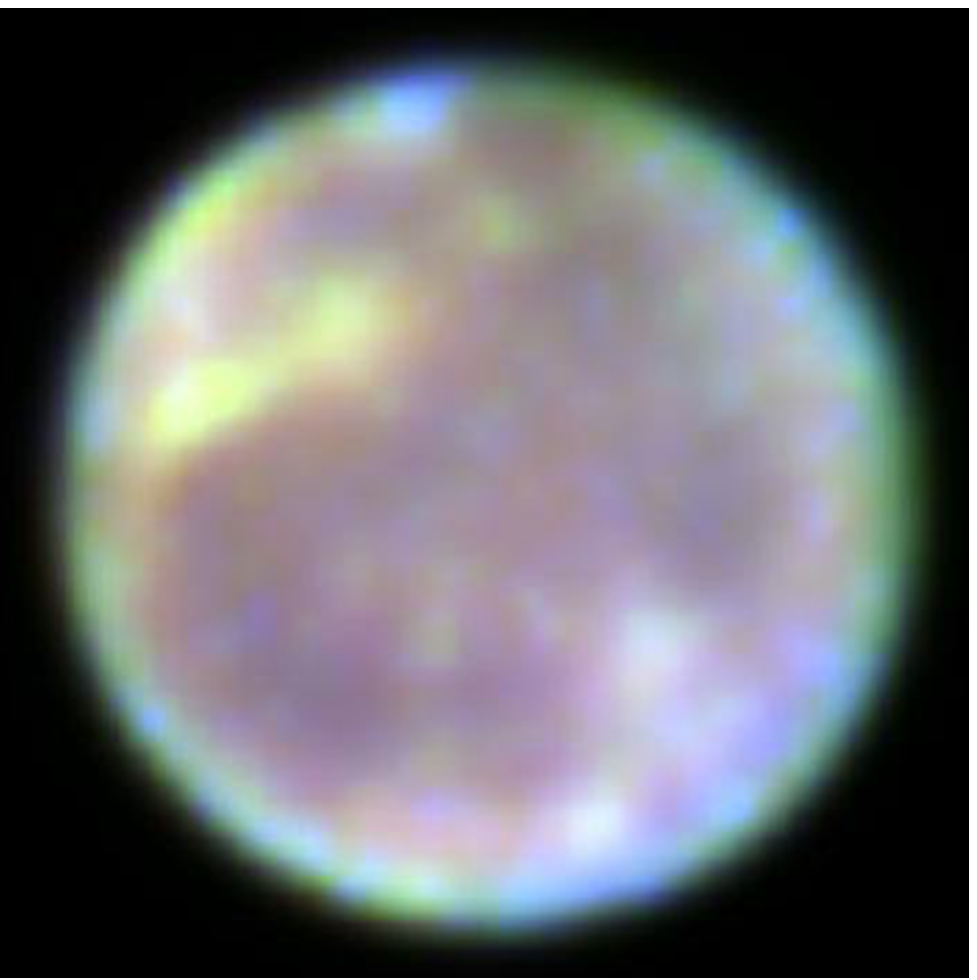}
% reference could be News Release Number: STSci-1995-35 (October 10, 1995).
\caption{(Left) Johnson-Cousins BRI false-color image of Ganymede, obtained with PALM-3000 and TMAS, demonstrating the visible-light correctional capability of the AO system.  The pixel sampling in this image is a mere 10 milliarcseconds, corresponding to about 35 km at the distance of Ganymede.  (Right) Hubble Space Telescope false-color image of Ganymede for comparison (NASA image).}
\label{Ganymede} 
\end{figure*}

An example of this potential is shown in Figure \ref{Ganymede}, where one hundred 0.5-second exposure images taken on 27 September 2012 of Ganymede taken in each of B, R, and I filters have been combined (Hildebrandt et al. 2013, in preparation), without  frame selection, into a false-color visual image following wavelet high-pass filter transformation.  In particular, solar system bodies subject to resurfacing, such as Io, Titan, and Triton, may in the future be periodically monitored at this resolution using PALM-3000.

\subsection{Future Work}
We have reported upon the initial on-sky demonstration of the $s_{64}$ subaperture correction mode of PALM-3000. Performance optimization in this mode to achieve wavefront errors as low as $\sigma$ = 105~nm RMS in median seeing are currently underway. Additional work will be required to implement the $s_{32}$ and $s_8$ pupil sampling modes, while implementation of $s_{16}$ operation has been indefinitely deferred by current funding limitations. Support of routine science operations has been fully transferred to Palomar Observatory staff, though engineering optimization for performance and observing efficiency continues through successive releases of additional AO system automation and calibration software.  

To date, all published PALM-3000 results have been based on the strategy of phase conjugation, which attempts to minimize the delivered science wavefront phase error.  Optimized performance for exoplanet contrast, however, will require implementation of full electric field (phase and amplitude) correction.  We have implemented both speckle nulling \citep{2006ApJ...638..488B} and Electric Field Conjugation \citep{2007SPIE.6691E...7G} algorithms for laboratory testing, but on-sky demonstration of these techniques with PALM-3000 remains a topic for future publication.

\section{Conclusions}
PALM-3000 is the key capability at the heart of a multi-dimensional science emphasis on exoplanet discovery and analysis.  Used in combination with the P1640 speckle-suppression coronagraph, PALM-3000 is the first realized AO system designed to achieve $10^{-7}$ contrast direct imaging and low-resolution near-infrared spectroscopy of exoplanets using a multi-layered approach of high-order adaptive optics, a spatially filtered Shack-Hartmann wavefront sensor, an apodized-pupil Lyot coronagraph, a speckle suppression integral field spectrograph, and nm-level accuracy calibration interferometer producing a dark hole in the stellar speckle field on sky.  Our Northern survey of A- and F-type stars in now underway.  When combined with small IWA = 0\farcs1 coronagraphs within PHARO and TMAS, PALM-3000 will also probe star-forming associations at distances of 150-300 pc at more modest, yet still unprecedented contrast levels.  With outer working angle as large a 3\farcs2, PALM-3000 will also probe a unique long-period orbital parameter space for exoplanets around nearby stars, including long-duration radial velocity trend stars identified at Keck Observatory and elsewhere.

As the highest actuator count adaptive optics system built to date for astronomical or, to our knowledge, any other purpose PALM-3000 is also being used for visible light science at spatial resolutions as fine as 16~mas for studies of several solar system bodies and circumstellar material around bright nearby stars.  Finally, PALM-3000 is successfully demonstrating technical solutions at the scale of deformable mirror and RTC capability required for the coming generation of 25-40~m diameter telescopes.

%Also will be a useful technique for not only detection of faint, massive planets, but also monitoring very close binaries \citep{hmo10}.

% If you wish to include an acknowledgments section in your paper,
% separate it off from the body of the text using the \acknowledgments
% command.
% Included in this acknowledgments section are examples of the
% AASTeX hypertext markup commands. Use \url without the optional [HREF]
% argument when you want to print the url directly in the text. Otherwise,
% use either \url or \anchor, with the HREF as the first argument and the
% text to be printed in the second.

\acknowledgments
This work was performed with financial support of National Science Foundation (NSF) through awards AST-0619922 and AST-1007046, Jet Propulsion Laboratory, Caltech Optical Observatories, and the generous philanthropy of Ron and Glo Helin.  Development of the Xinetics, Inc. The PALM-3000 HODM was funded by NASA SBIR award \#NNG05CA21C.  The successful deployment of PALM-3000 could not have been possible without the notable talent and dedication of the entire Palomar Observatory staff.  We gratefully acknowledge the specific contributions from Dan McKenna, John Henning, Steve Kunsman, Kajsa Peffer, Jean Mueller, Kevin Rykowski, Carolyn Heffner, Steve Einer, Greg van Idsinga, Mike Doyle, and Bruce Baker in interfacing to the Hale Telescope and to obtaining the commissioning data described herein.

{\it Facilities:} \facility{Hale(PALM-3000)}, \facility{Hale(P1640)}, \facility{Hale(TMAS)}, \facility{Hale(PHARO)}

%\appendix 

% The reference list follows the main body and any appendices.
% Use LaTeX's thebibliography environment to mark up your reference list.
% Note \begin{thebibliography} is followed by an empty set of
% curly braces. If you forget this, LaTeX will generate the error
% "Perhaps a missing \item?".
%
% thebibliography produces citations in the text using \bibitem-\cite
% cross-referencing. Each reference is preceded by a
% \bibitem command that defines in curly braces the KEY that corresponds
% to the KEY in the \cite commands (see the first section above).
% Make sure that you provide a unique KEY for every \bibitem or else the
% paper will not LaTeX. The square brackets should contain
% the citation text that LaTeX will insert in
% place of the \cite commands.
% We have used macros to produce journal name abbreviations.
% AASTeX provides a number of these for the more frequently-cited journals.
% See the Author Guide for a list of them.
% Note that the style of the \bibitem labels (in []) is slightly
% different from previous examples. The natbib system solves a host
% of citation expression problems, but it is necessary to clearly
% delimit the year from the author name used in the citation.
% See the natbib documentation for more details and options.

\bibliographystyle{plainnat}
%\bibliography{/Users/Richard Dekany/COO/Projects/PALM-3000/Papers/ao_v2}   %>>>> bibliography data in report.bib
%\bibliography{ao_v2}   %>>>> bibliography data in report.bib

\begin{thebibliography}{84}
\providecommand{\natexlab}[1]{#1}
\providecommand{\url}[1]{\texttt{#1}}
\expandafter\ifx\csname urlstyle\endcsname\relax
  \providecommand{\doi}[1]{doi: #1}\else
  \providecommand{\doi}{doi: \begingroup \urlstyle{rm}\Url}\fi

\bibitem[{Angel}(1994)]{1994Natur.368..203A}
J.~R.~P. {Angel}.
\newblock {Ground-based imaging of extrasolar planets using adaptive optics}.
\newblock \emph{\nat}, 368:\penalty0 203--207, March 1994.
\newblock \doi{10.1038/368203a0}.

\bibitem[{Asensio Ramos} and {L{\'o}pez Ariste}(2010)]{2010A&A...518A...6A}
A.~{Asensio Ramos} and A.~{L{\'o}pez Ariste}.
\newblock {Image reconstruction with analytical point spread functions}.
\newblock \emph{\aap}, 518:\penalty0 A6, July 2010.
\newblock \doi{10.1051/0004-6361/201014318}.

\bibitem[{Baranec}(2008)]{2008SPIE.7015E.152B}
C.~{Baranec}.
\newblock {High-order wavefront sensing system for PALM-3000}.
\newblock In \emph{Society of Photo-Optical Instrumentation Engineers (SPIE)
  Conference Series}, volume 7015 of \emph{Society of Photo-Optical
  Instrumentation Engineers (SPIE) Conference Series}, July 2008.
\newblock \doi{10.1117/12.788460}.

\bibitem[Baranec and Dekany(2008)]{Baranec:08}
Christoph Baranec and Richard Dekany.
\newblock Study of a mems-based shack-hartmann wavefront sensor with adjustable
  pupil sampling for astronomical adaptive optics.
\newblock \emph{Appl. Opt.}, 47\penalty0 (28):\penalty0 5155--5162, Oct 2008.
\newblock \doi{10.1364/AO.47.005155}.
\newblock URL \url{http://ao.osa.org/abstract.cfm?URI=ao-47-28-5155}.

\bibitem[{Beichman} et~al.(2010){Beichman}, {Krist}, {Trauger}, {Greene},
  {Oppenheimer}, {Sivaramakrishnan}, {Doyon}, {Boccaletti}, {Barman}, and
  {Rieke}]{2010PASP..122..162B}
C.~A. {Beichman}, J.~{Krist}, J.~T. {Trauger}, T.~{Greene}, B.~{Oppenheimer},
  A.~{Sivaramakrishnan}, R.~{Doyon}, A.~{Boccaletti}, T.~S. {Barman}, and
  M.~{Rieke}.
\newblock {Imaging Young Giant Planets From Ground and Space}.
\newblock \emph{\pasp}, 122:\penalty0 162--200, February 2010.
\newblock \doi{10.1086/651057}.

\bibitem[{Bernat} et~al.(2010){Bernat}, {Bouchez}, {Ireland}, {Tuthill},
  {Martinache}, {Angione}, {Burruss}, {Cromer}, {Dekany}, {Guiwits}, {Henning},
  {Hickey}, {Kibblewhite}, {McKenna}, {Moore}, {Petrie}, {Roberts}, {Shelton},
  {Thicksten}, {Trinh}, {Tripathi}, {Troy}, {Truong}, {Velur}, and
  {Lloyd}]{2010ApJ...715..724B}
D.~{Bernat}, A.~H. {Bouchez}, M.~{Ireland}, P.~{Tuthill}, F.~{Martinache},
  J.~{Angione}, R.~S. {Burruss}, J.~L. {Cromer}, R.~G. {Dekany}, S.~R.
  {Guiwits}, J.~R. {Henning}, J.~{Hickey}, E.~{Kibblewhite}, D.~L. {McKenna},
  A.~M. {Moore}, H.~L. {Petrie}, J.~{Roberts}, J.~C. {Shelton}, R.~P.
  {Thicksten}, T.~{Trinh}, R.~{Tripathi}, M.~{Troy}, T.~{Truong}, V.~{Velur},
  and J.~P. {Lloyd}.
\newblock {A Close Companion Search Around L Dwarfs Using Aperture Masking
  Interferometry and Palomar Laser Guide Star Adaptive Optics}.
\newblock \emph{\apj}, 715:\penalty0 724--735, June 2010.
\newblock \doi{10.1088/0004-637X/715/2/724}.

\bibitem[{Bloemhof} et~al.(2000{\natexlab{a}}){Bloemhof}, {Marsh}, {Dekany},
  {Troy}, {Marshall}, {Oppenheimer}, {Hayward}, and
  {Brandl}]{2000SPIE.4007..889B}
E.~E. {Bloemhof}, K.~A. {Marsh}, R.~G. {Dekany}, M.~{Troy}, J.~{Marshall},
  B.~R. {Oppenheimer}, T.~L. {Hayward}, and B.~{Brandl}.
\newblock {Stability of the adaptive-optic point spread function: metrics,
  deconvolution, and initial Palomar results}.
\newblock In {P.~L.~Wizinowich}, editor, \emph{Society of Photo-Optical
  Instrumentation Engineers (SPIE) Conference Series}, volume 4007 of
  \emph{Society of Photo-Optical Instrumentation Engineers (SPIE) Conference
  Series}, pages 889--898, July 2000{\natexlab{a}}.

\bibitem[{Bloemhof} et~al.(2000{\natexlab{b}}){Bloemhof}, {Oppenheimer},
  {Dekany}, {Troy}, {Hayward}, and {Brandl}]{2000SPIE.4007..839B}
E.~E. {Bloemhof}, B.~R. {Oppenheimer}, R.~G. {Dekany}, M.~{Troy}, T.~L.
  {Hayward}, and B.~{Brandl}.
\newblock {Studies of Herbig-Haro objects with the Palomar adaptive optics
  system}.
\newblock In {P.~L.~Wizinowich}, editor, \emph{Society of Photo-Optical
  Instrumentation Engineers (SPIE) Conference Series}, volume 4007 of
  \emph{Society of Photo-Optical Instrumentation Engineers (SPIE) Conference
  Series}, pages 839--846, July 2000{\natexlab{b}}.

\bibitem[{Bloemhof} et~al.(2001){Bloemhof}, {Dekany}, {Troy}, and
  {Oppenheimer}]{2001ApJ...558L..71B}
E.~E. {Bloemhof}, R.~G. {Dekany}, M.~{Troy}, and B.~R. {Oppenheimer}.
\newblock {Behavior of Remnant Speckles in an Adaptively Corrected Imaging
  System}.
\newblock \emph{\apjl}, 558:\penalty0 L71--L74, September 2001.
\newblock \doi{10.1086/323494}.

\bibitem[{Boccaletti} et~al.(2003){Boccaletti}, {Chauvin}, {Lagrange}, and
  {Marchis}]{2003A&A...410..283B}
A.~{Boccaletti}, G.~{Chauvin}, A.-M. {Lagrange}, and F.~{Marchis}.
\newblock {Near-IR coronagraphic imaging of the companion to HR 7672}.
\newblock \emph{\aap}, 410:\penalty0 283--288, October 2003.
\newblock \doi{10.1051/0004-6361:20031216}.

\bibitem[{Bord{\'e}} and {Traub}(2006)]{2006ApJ...638..488B}
P.~J. {Bord{\'e}} and W.~A. {Traub}.
\newblock {High-Contrast Imaging from Space: Speckle Nulling in a
  Low-Aberration Regime}.
\newblock \emph{\apj}, 638:\penalty0 488--498, February 2006.
\newblock \doi{10.1086/498669}.

\bibitem[{Bouchez} et~al.(2007){Bouchez}, {Dekany}, {Angione}, {Burruss},
  {Cromer}, {Guiwits}, {Henning}, {Hickey}, {Kibblewhite}, {Moore}, {Petrie},
  {Roberts}, {Shelton}, {Thicksten}, {Trinh}, {Tripathi}, {Troy}, {Truong}, and
  {Velur}]{2007AAS...211.5803B}
A.~H. {Bouchez}, R.~G. {Dekany}, J.~{Angione}, R.~{Burruss}, J.~{Cromer},
  S.~{Guiwits}, J.~R. {Henning}, J.~{Hickey}, E.~{Kibblewhite}, A.~{Moore},
  H.~L. {Petrie}, J.~{Roberts}, J.~C. {Shelton}, R.~P. {Thicksten}, T.~{Trinh},
  R.~{Tripathi}, M.~{Troy}, T.~{Truong}, and V.~{Velur}.
\newblock {Palomar Laser Guide Star Adaptive Optics Observations of Globular
  Cluster GLIMPSE-C01}.
\newblock In \emph{American Astronomical Society Meeting Abstracts}, volume~38
  of \emph{Bulletin of the American Astronomical Society}, page 834, December
  2007.

\bibitem[{Brennan} and {Kim}(2001)]{2001JSV...246..281B}
M.~J. {Brennan} and S.-M. {Kim}.
\newblock {Feedforward and Feedback Control of Sound and Vibration-A Wiener
  Filter Approach}.
\newblock \emph{Journal of Sound Vibration}, 246:\penalty0 281--296, September
  2001.

\bibitem[{Burruss} et~al.(2010){Burruss}, {Serabyn}, {Mawet}, {Roberts},
  {Hickey}, {Rykoski}, {Bikkannavar}, and {Crepp}]{2010SPIE.7736E.197B}
R.~S. {Burruss}, E.~{Serabyn}, D.~P. {Mawet}, J.~E. {Roberts}, J.~P. {Hickey},
  K.~{Rykoski}, S.~{Bikkannavar}, and J.~R. {Crepp}.
\newblock {Demonstration of on sky contrast improvement using the modified
  Gerchberg-Saxton algorithm at the Palomar Observatory}.
\newblock In \emph{Society of Photo-Optical Instrumentation Engineers (SPIE)
  Conference Series}, volume 7736 of \emph{Society of Photo-Optical
  Instrumentation Engineers (SPIE) Conference Series}, July 2010.
\newblock \doi{10.1117/12.857544}.

\bibitem[{Chauvin} et~al.(2012){Chauvin}, {Lagrange}, {Beust}, {Bonnefoy},
  {Boccaletti}, {Apai}, {Allard}, {Ehrenreich}, {Girard}, {Mouillet}, and
  {Rouan}]{2012A&A...542A..41C}
G.~{Chauvin}, A.-M. {Lagrange}, H.~{Beust}, M.~{Bonnefoy}, A.~{Boccaletti},
  D.~{Apai}, F.~{Allard}, D.~{Ehrenreich}, J.~H.~V. {Girard}, D.~{Mouillet},
  and D.~{Rouan}.
\newblock {Orbital characterization of the {$\beta$} Pictoris b giant planet}.
\newblock \emph{\aap}, 542:\penalty0 A41, June 2012.
\newblock \doi{10.1051/0004-6361/201118346}.

\bibitem[{Chiang} et~al.(2009){Chiang}, {Kite}, {Kalas}, {Graham}, and
  {Clampin}]{2009ApJ...693..734C}
E.~{Chiang}, E.~{Kite}, P.~{Kalas}, J.~R. {Graham}, and M.~{Clampin}.
\newblock {Fomalhaut's Debris Disk and Planet: Constraining the Mass of
  Fomalhaut b from disk Morphology}.
\newblock \emph{\apj}, 693:\penalty0 734--749, March 2009.
\newblock \doi{10.1088/0004-637X/693/1/734}.

\bibitem[{Crepp} and {Johnson}(2011)]{2011ApJ...733..126C}
J.~R. {Crepp} and J.~A. {Johnson}.
\newblock {Estimates of the Planet Yield from Ground-based High-contrast
  Imaging Observations as a Function of Stellar Mass}.
\newblock \emph{\apj}, 733:\penalty0 126, June 2011.
\newblock \doi{10.1088/0004-637X/733/2/126}.

\bibitem[{Davies} and {Kasper}(2012)]{2012ARA&A..50..305D}
R.~{Davies} and M.~{Kasper}.
\newblock {Adaptive Optics for Astronomy}.
\newblock \emph{\araa}, 50:\penalty0 305--351, September 2012.
\newblock \doi{10.1146/annurev-astro-081811-125447}.

\bibitem[{Dekany} et~al.(2006){Dekany}, {Bouchez}, {Britton}, {Velur}, {Troy},
  {Shelton}, and {Roberts}]{2006SPIE.6272E..13D}
R.~{Dekany}, A.~{Bouchez}, M.~{Britton}, V.~{Velur}, M.~{Troy}, J.~C.
  {Shelton}, and J.~{Roberts}.
\newblock {PALM-3000: visible light AO on the 5.1-meter Telescope}.
\newblock In \emph{Society of Photo-Optical Instrumentation Engineers (SPIE)
  Conference Series}, volume 6272 of \emph{Society of Photo-Optical
  Instrumentation Engineers (SPIE) Conference Series}, July 2006.
\newblock \doi{10.1117/12.674496}.

\bibitem[{Dekany} et~al.(1997){Dekany}, {Wallace}, {Brack}, {Oppenheimer}, and
  {Palmer}]{1997SPIE.3126..269D}
R.~G. {Dekany}, J.~K. {Wallace}, G.~{Brack}, B.~R. {Oppenheimer}, and
  D.~{Palmer}.
\newblock {Initial test results from the Palomar 200-in. adaptive optics system
  [3126-33]}.
\newblock In {R.~K.~Tyson \& R.~Q.~Fugate}, editor, \emph{Society of
  Photo-Optical Instrumentation Engineers (SPIE) Conference Series}, volume
  3126 of \emph{Society of Photo-Optical Instrumentation Engineers (SPIE)
  Conference Series}, page 269, October 1997.
\newblock \doi{10.1117/12.290153}.

\bibitem[{Dekany} et~al.(2003){Dekany}, {Bauman}, {Gavel}, {Troy}, {Macintosh},
  and {Britton}]{2003SPIE.4839.1165D}
R.~G. {Dekany}, B.~J. {Bauman}, D.~T. {Gavel}, M.~{Troy}, B.~A. {Macintosh},
  and M.~C. {Britton}.
\newblock {Initial concepts for CELT adaptive optics}.
\newblock In {P.~L.~Wizinowich \& D.~Bonaccini}, editor, \emph{Society of
  Photo-Optical Instrumentation Engineers (SPIE) Conference Series}, volume
  4839 of \emph{Society of Photo-Optical Instrumentation Engineers (SPIE)
  Conference Series}, pages 1165--1174, February 2003.
\newblock \doi{10.1117/12.459798}.

\bibitem[Dekany and Holm(1989)]{Dekany89}
Richard Dekany and Ron Holm.
\newblock Conic and nested cone retroreflectors.
\newblock 18005/7502, 1989.

\bibitem[{DuVarney} et~al.(2000{\natexlab{a}}){DuVarney}, {Bleau}, {Motter},
  {Dekany}, {Troy}, and {Brack}]{2000ASSL..252..395D}
R.~{DuVarney}, C.~{Bleau}, G.~{Motter}, R.~{Dekany}, M.~{Troy}, and G.~{Brack}.
\newblock {SciMeasure wavefront sensor cameras and their application in the
  Palomar adaptive optics system}.
\newblock In {P.~Amico \& J.~W.~Beletic}, editor, \emph{Astrophysics and Space
  Science Library}, volume 252 of \emph{Astrophysics and Space Science
  Library}, page 395, 2000{\natexlab{a}}.

\bibitem[{DuVarney} et~al.(2001){DuVarney}, {Bleau}, {Motter}, {Dekany},
  {Troy}, and {Brack}]{2001ExA....11..237D}
R.~{DuVarney}, C.~{Bleau}, G.~{Motter}, R.~{Dekany}, M.~{Troy}, and G.~{Brack}.
\newblock {SciMeasure Wavefront Sensor Cameras and their Application in the
  Palomar Adaptive Optics System}.
\newblock \emph{Experimental Astronomy}, 11:\penalty0 237--249, 2001.
\newblock \doi{10.1023/A:1013199214825}.

\bibitem[{DuVarney} et~al.(2000{\natexlab{b}}){DuVarney}, {Bleau}, {Motter},
  {Shaklan}, {Kuhnert}, {Brack}, {Palmer}, {Troy}, {Kieu}, and
  {Dekany}]{2000SPIE.4007..481D}
R.~C. {DuVarney}, C.~A. {Bleau}, G.~T. {Motter}, S.~B. {Shaklan}, A.~C.
  {Kuhnert}, G.~{Brack}, D.~{Palmer}, M.~{Troy}, T.~{Kieu}, and R.~G. {Dekany}.
\newblock {EEV CCD39 wavefront sensor cameras for AO and interferometry}.
\newblock In {P.~L.~Wizinowich}, editor, \emph{Society of Photo-Optical
  Instrumentation Engineers (SPIE) Conference Series}, volume 4007 of
  \emph{Society of Photo-Optical Instrumentation Engineers (SPIE) Conference
  Series}, pages 481--492, July 2000{\natexlab{b}}.

\bibitem[Femen{\'\i}a~Castell{\'a} et~al.(2010)Femen{\'\i}a~Castell{\'a},
  Labadie, Rebolo~L{\'o}pez, P{\'e}rez~Prieto, P{\'e}rez~Garrido,
  D{\'\i}az~S{\'a}nchez, and Villo~P{\'e}rez]{doi:10.1117/12.857245}
B.~Femen{\'\i}a~Castell{\'a}, L.~Labadie, R.~Rebolo~L{\'o}pez, J.~A.
  P{\'e}rez~Prieto, A.~P{\'e}rez~Garrido, A.~D{\'\i}az~S{\'a}nchez, and
  I.~Villo~P{\'e}rez.
\newblock Lucky imaging and adaptive optics on 10-m class telescopes: a real
  promise for diffraction limited imaging in the visible?
\newblock pages 77363X--77363X--11, 2010.
\newblock \doi{10.1117/12.857245}.
\newblock URL \url{+ http://dx.doi.org/10.1117/12.857245}.

\bibitem[Fried(1977)]{Fried:77}
David~L. Fried.
\newblock Least-square fitting a wave-front distortion estimate to an array of
  phase-difference measurements.
\newblock \emph{J. Opt. Soc. Am.}, 67\penalty0 (3):\penalty0 370--375, Mar
  1977.
\newblock \doi{10.1364/JOSA.67.000370}.
\newblock URL
  \url{http://www.opticsinfobase.org/abstract.cfm?URI=josa-67-3-370}.

\bibitem[Garrel et~al.(2011)Garrel, Guyon, Baudoz, Martinache, Stewart, Lozi,
  and Groff]{doi:10.1117/12.894309}
Vincent Garrel, Olivier Guyon, Pierre Baudoz, Frantz Martinache, Paul Stewart,
  Julien Lozi, and Tyler Groff.
\newblock The subaru coronagraphic extreme ao (scexao) system: fast visible
  imager.
\newblock pages 81510R--81510R--8, 2011.

\bibitem[Geldard(1972)]{Geldard72}
F.~A. Geldard.
\newblock \emph{The Human Senses}.
\newblock 1972.

\bibitem[{Give'on} et~al.(2007){Give'on}, {Kern}, {Shaklan}, {Moody}, and
  {Pueyo}]{2007SPIE.6691E...7G}
A.~{Give'on}, B.~{Kern}, S.~{Shaklan}, D.~C. {Moody}, and L.~{Pueyo}.
\newblock {Broadband wavefront correction algorithm for high-contrast imaging
  systems}.
\newblock In \emph{Society of Photo-Optical Instrumentation Engineers (SPIE)
  Conference Series}, volume 6691 of \emph{Society of Photo-Optical
  Instrumentation Engineers (SPIE) Conference Series}, September 2007.
\newblock \doi{10.1117/12.733122}.

\bibitem[{Gonz{\'a}lez-Garc{\'{\i}}a} et~al.(2006){Gonz{\'a}lez-Garc{\'{\i}}a},
  {Zapatero Osorio}, {B{\'e}jar}, {Bihain}, {Barrado Y Navascu{\'e}s},
  {Caballero}, and {Morales-Calder{\'o}n}]{2006A&A...460..799G}
B.~M. {Gonz{\'a}lez-Garc{\'{\i}}a}, M.~R. {Zapatero Osorio}, V.~J.~S.
  {B{\'e}jar}, G.~{Bihain}, D.~{Barrado Y Navascu{\'e}s}, J.~A. {Caballero},
  and M.~{Morales-Calder{\'o}n}.
\newblock {A search for substellar members in the Praesepe and {$\sigma$}
  Orionis clusters}.
\newblock \emph{\aap}, 460:\penalty0 799--810, December 2006.
\newblock \doi{10.1051/0004-6361:20065909}.

\bibitem[{Haisch} et~al.(2001){Haisch}, {Lada}, and
  {Lada}]{2001ApJ...553L.153H}
K.~E. {Haisch}, Jr., E.~A. {Lada}, and C.~J. {Lada}.
\newblock {Disk Frequencies and Lifetimes in Young Clusters}.
\newblock \emph{\apjl}, 553:\penalty0 L153--L156, June 2001.
\newblock \doi{10.1086/320685}.

\bibitem[{Hanot} et~al.(2011){Hanot}, {Mennesson}, {Martin}, {Liewer}, {Loya},
  {Mawet}, {Riaud}, {Absil}, and {Serabyn}]{2011ApJ...729..110H}
C.~{Hanot}, B.~{Mennesson}, S.~{Martin}, K.~{Liewer}, F.~{Loya}, D.~{Mawet},
  P.~{Riaud}, O.~{Absil}, and E.~{Serabyn}.
\newblock {Improving Interferometric Null Depth Measurements using Statistical
  Distributions: Theory and First Results with the Palomar Fiber Nuller}.
\newblock \emph{\apj}, 729:\penalty0 110, March 2011.
\newblock \doi{10.1088/0004-637X/729/2/110}.

\bibitem[Hardy(1998)]{Hardy98}
John~W. Hardy.
\newblock \emph{Adaptive Optics for Astronomical Telescopes}.
\newblock Oxford University Press, New York, N.Y., 1998.

\bibitem[Hart(2010)]{Hart:10}
Michael Hart.
\newblock Recent advances in astronomical adaptive optics.
\newblock \emph{Appl. Opt.}, 49\penalty0 (16):\penalty0 D17--D29, Jun 2010.
\newblock \doi{10.1364/AO.49.000D17}.
\newblock URL \url{http://ao.osa.org/abstract.cfm?URI=ao-49-16-D17}.

\bibitem[{Haubois} et~al.(2009){Haubois}, {Perrin}, {Lacour}, {Verhoelst},
  {Meimon}, {Mugnier}, {Thi{\'e}baut}, {Berger}, {Ridgway}, {Monnier},
  {Millan-Gabet}, and {Traub}]{2009A&A...508..923H}
X.~{Haubois}, G.~{Perrin}, S.~{Lacour}, T.~{Verhoelst}, S.~{Meimon},
  L.~{Mugnier}, E.~{Thi{\'e}baut}, J.~P. {Berger}, S.~T. {Ridgway}, J.~D.
  {Monnier}, R.~{Millan-Gabet}, and W.~{Traub}.
\newblock {Imaging the spotty surface of <ASTROBJ>Betelgeuse</ASTROBJ> in the H
  band}.
\newblock \emph{\aap}, 508:\penalty0 923--932, December 2009.
\newblock \doi{10.1051/0004-6361/200912927}.

\bibitem[{Hayward} et~al.(2001){Hayward}, {Brandl}, {Pirger}, {Blacken},
  {Gull}, {Schoenwald}, and {Houck}]{2001PASP..113..105H}
T.~L. {Hayward}, B.~{Brandl}, B.~{Pirger}, C.~{Blacken}, G.~E. {Gull},
  J.~{Schoenwald}, and J.~R. {Houck}.
\newblock {PHARO: A Near-Infrared Camera for the Palomar Adaptive Optics
  System}.
\newblock \emph{\pasp}, 113:\penalty0 105--118, January 2001.
\newblock \doi{10.1086/317969}.

\bibitem[{Hinkley} et~al.(2011){Hinkley}, {Oppenheimer}, {Zimmerman},
  {Brenner}, {Parry}, {Crepp}, {Vasisht}, {Ligon}, {King}, {Soummer},
  {Sivaramakrishnan}, {Beichman}, {Shao}, {Roberts}, {Bouchez}, {Dekany},
  {Pueyo}, {Roberts}, {Lockhart}, {Zhai}, {Shelton}, and
  {Burruss}]{2011PASP..123...74H}
S.~{Hinkley}, B.~R. {Oppenheimer}, N.~{Zimmerman}, D.~{Brenner}, I.~R. {Parry},
  J.~R. {Crepp}, G.~{Vasisht}, E.~{Ligon}, D.~{King}, R.~{Soummer},
  A.~{Sivaramakrishnan}, C.~{Beichman}, M.~{Shao}, L.~C. {Roberts},
  A.~{Bouchez}, R.~{Dekany}, L.~{Pueyo}, J.~E. {Roberts}, T.~{Lockhart},
  C.~{Zhai}, C.~{Shelton}, and R.~{Burruss}.
\newblock {A New High Contrast Imaging Program at Palomar Observatory}.
\newblock \emph{\pasp}, 123:\penalty0 74--86, January 2011.
\newblock \doi{10.1086/658163}.

\bibitem[{Johnson} et~al.(2007){Johnson}, {Butler}, {Marcy}, {Fischer}, {Vogt},
  {Wright}, and {Peek}]{2007ApJ...670..833J}
J.~A. {Johnson}, R.~P. {Butler}, G.~W. {Marcy}, D.~A. {Fischer}, S.~S. {Vogt},
  J.~T. {Wright}, and K.~M.~G. {Peek}.
\newblock {A New Planet around an M Dwarf: Revealing a Correlation between
  Exoplanets and Stellar Mass}.
\newblock \emph{\apj}, 670:\penalty0 833--840, November 2007.
\newblock \doi{10.1086/521720}.

\bibitem[{Jolissaint} et~al.(2006){Jolissaint}, {V{\'e}ran}, and
  {Conan}]{2006JOSAA..23..382J}
L.~{Jolissaint}, J.-P. {V{\'e}ran}, and R.~{Conan}.
\newblock {Analytical modeling of adaptive optics: foundations of the phase
  spatial power spectrum approach}.
\newblock \emph{Journal of the Optical Society of America A}, 23:\penalty0
  382--394, February 2006.
\newblock \doi{10.1364/JOSAA.23.000382}.

\bibitem[{Kalas} et~al.(2008){Kalas}, {Graham}, {Chiang}, {Fitzgerald},
  {Clampin}, {Kite}, {Stapelfeldt}, {Marois}, and {Krist}]{2008Sci...322.1345K}
P.~{Kalas}, J.~R. {Graham}, E.~{Chiang}, M.~P. {Fitzgerald}, M.~{Clampin},
  E.~S. {Kite}, K.~{Stapelfeldt}, C.~{Marois}, and J.~{Krist}.
\newblock {Optical Images of an Exosolar Planet 25 Light-Years from Earth}.
\newblock \emph{Science}, 322:\penalty0 1345--, November 2008.
\newblock \doi{10.1126/science.1166609}.

\bibitem[{Kibblewhite}(2008)]{2008SPIE.7015E..14K}
E.~{Kibblewhite}.
\newblock {Calculation of returns from sodium beacons for different types of
  laser}.
\newblock In \emph{Society of Photo-Optical Instrumentation Engineers (SPIE)
  Conference Series}, volume 7015 of \emph{Society of Photo-Optical
  Instrumentation Engineers (SPIE) Conference Series}, July 2008.
\newblock \doi{10.1117/12.789601}.

\bibitem[{Kibblewhite}(2009)]{2009amos.confE..33K}
E.~{Kibblewhite}.
\newblock {The Physics of the SODIUM Laser Guide Stat: Predicting and Enhancing
  the Photon Returns of Sodium Guide Stars for Different Laser Technologies}.
\newblock In \emph{Advanced Maui Optical and Space Surveillance Technologies
  Conference,}, 2009.

\bibitem[{Kraus} et~al.(2008){Kraus}, {Ireland}, {Martinache}, and
  {Lloyd}]{2008ApJ...679..762K}
A.~L. {Kraus}, M.~J. {Ireland}, F.~{Martinache}, and J.~P. {Lloyd}.
\newblock {Mapping the Shores of the Brown Dwarf Desert. I. Upper Scorpius}.
\newblock \emph{\apj}, 679:\penalty0 762--782, May 2008.
\newblock \doi{10.1086/587435}.

\bibitem[{Lagadec} et~al.(2011){Lagadec}, {Verhoelst}, {M{\'e}karnia},
  {Su{\'a}eez}, {Zijlstra}, {Bendjoya}, {Szczerba}, {Chesneau}, {van Winckel},
  {Barlow}, {Matsuura}, {Bowey}, {Lorenz-Martins}, and
  {Gledhill}]{2011MNRAS.417...32L}
E.~{Lagadec}, T.~{Verhoelst}, D.~{M{\'e}karnia}, O.~{Su{\'a}eez}, A.~A.
  {Zijlstra}, P.~{Bendjoya}, R.~{Szczerba}, O.~{Chesneau}, H.~{van Winckel},
  M.~J. {Barlow}, M.~{Matsuura}, J.~E. {Bowey}, S.~{Lorenz-Martins}, and
  T.~{Gledhill}.
\newblock {A mid-infrared imaging catalogue of post-asymptotic giant branch
  stars}.
\newblock \emph{\mnras}, 417:\penalty0 32--92, October 2011.
\newblock \doi{10.1111/j.1365-2966.2011.18557.x}.

\bibitem[{Law} et~al.(2009){Law}, {Mackay}, {Dekany}, {Ireland}, {Lloyd},
  {Moore}, {Robertson}, {Tuthill}, and {Woodruff}]{2009ApJ...692..924L}
N.~M. {Law}, C.~D. {Mackay}, R.~G. {Dekany}, M.~{Ireland}, J.~P. {Lloyd}, A.~M.
  {Moore}, J.~G. {Robertson}, P.~{Tuthill}, and H.~C. {Woodruff}.
\newblock {Getting Lucky with Adaptive Optics: Fast Adaptive Optics Image
  Selection in the Visible with a Large Telescope}.
\newblock \emph{\apj}, 692:\penalty0 924--930, February 2009.
\newblock \doi{10.1088/0004-637X/692/1/924}.

\bibitem[Law and Lane(1996)]{Law:1996}
N.F Law and R.G Lane.
\newblock Wavefront estimation at low light levels.
\newblock \emph{Optics Communications}, 126\penalty0 (1--3):\penalty0 19 -- 24,
  1996.
\newblock ISSN 0030-4018.
\newblock \doi{10.1016/0030-4018(96)00027-2}.
\newblock URL
  \url{http://www.sciencedirect.com/science/article/pii/0030401896000272}.

\bibitem[{Lloyd} et~al.(2006){Lloyd}, {Martinache}, {Ireland}, {Monnier},
  {Pravdo}, {Shaklan}, and {Tuthill}]{2006ApJ...650L.131L}
J.~P. {Lloyd}, F.~{Martinache}, M.~J. {Ireland}, J.~D. {Monnier}, S.~H.
  {Pravdo}, S.~B. {Shaklan}, and P.~G. {Tuthill}.
\newblock {Direct Detection of the Brown Dwarf GJ 802B with Adaptive Optics
  Masking Interferometry}.
\newblock \emph{\apjl}, 650:\penalty0 L131--L134, October 2006.
\newblock \doi{10.1086/508771}.

\bibitem[MacMartin(2003)]{MacMartin:03}
Douglas~G. MacMartin.
\newblock Local, hierarchic, and iterative reconstructors for adaptive optics.
\newblock \emph{J. Opt. Soc. Am. A}, 20\penalty0 (6):\penalty0 1084--1093, Jun
  2003.
\newblock \doi{10.1364/JOSAA.20.001084}.
\newblock URL \url{http://josaa.osa.org/abstract.cfm?URI=josaa-20-6-1084}.

\bibitem[Marechal(1947)]{Marechal47}
A.~Marechal.
\newblock \emph{Rev d'Optique}, 26:\penalty0 257, 1947.

\bibitem[{Marois} et~al.(2008){Marois}, {Macintosh}, {Barman}, {Zuckerman},
  {Song}, {Patience}, {Lafreni{\`e}re}, and {Doyon}]{2008Sci...322.1348M}
C.~{Marois}, B.~{Macintosh}, T.~{Barman}, B.~{Zuckerman}, I.~{Song},
  J.~{Patience}, D.~{Lafreni{\`e}re}, and R.~{Doyon}.
\newblock {Direct Imaging of Multiple Planets Orbiting the Star HR 8799}.
\newblock \emph{Science}, 322:\penalty0 1348--, November 2008.
\newblock \doi{10.1126/science.1166585}.

\bibitem[{Mawet} et~al.(2012){Mawet}, {Pueyo}, {Lawson}, {Mugnier}, {Traub},
  {Boccaletti}, {Trauger}, {Gladysz}, {Serabyn}, {Milli}, {Belikov}, {Kasper},
  {Baudoz}, {Macintosh}, {Marois}, {Oppenheimer}, {Barrett}, {Beuzit},
  {Devaney}, {Girard}, {Guyon}, {Krist}, {Mennesson}, {Mouillet}, {Murakami},
  {Poyneer}, {Savransky}, {V{\'e}rinaud}, and {Wallace}]{2012SPIE.8442E..04M}
D.~{Mawet}, L.~{Pueyo}, P.~{Lawson}, L.~{Mugnier}, W.~{Traub}, A.~{Boccaletti},
  J.~T. {Trauger}, S.~{Gladysz}, E.~{Serabyn}, J.~{Milli}, R.~{Belikov},
  M.~{Kasper}, P.~{Baudoz}, B.~{Macintosh}, C.~{Marois}, B.~{Oppenheimer},
  H.~{Barrett}, J.-L. {Beuzit}, N.~{Devaney}, J.~{Girard}, O.~{Guyon},
  J.~{Krist}, B.~{Mennesson}, D.~{Mouillet}, N.~{Murakami}, L.~{Poyneer},
  D.~{Savransky}, C.~{V{\'e}rinaud}, and J.~K. {Wallace}.
\newblock {Review of small-angle coronagraphic techniques in the wake of
  ground-based second-generation adaptive optics systems}.
\newblock In \emph{Society of Photo-Optical Instrumentation Engineers (SPIE)
  Conference Series}, volume 8442 of \emph{Society of Photo-Optical
  Instrumentation Engineers (SPIE) Conference Series}, September 2012.
\newblock \doi{10.1117/12.927245}.

\bibitem[{Mawet} et~al.(2013){Mawet}, {Absil}, {Delacroix}, {Girard}, {Milli},
  {O'Neal}, {Baudoz}, {Boccaletti}, {Bourget}, {Christiaens}, {Forsberg},
  {Gonte}, {Habraken}, {Hanot}, {Karlsson}, {Kasper}, {Lizon}, {Muzic},
  {Olivier}, {Pe{\~n}a}, {Slusarenko}, {Tacconi-Garman}, and
  {Surdej}]{2013A&A...552L..13M}
D.~{Mawet}, O.~{Absil}, C.~{Delacroix}, J.~H. {Girard}, J.~{Milli},
  J.~{O'Neal}, P.~{Baudoz}, A.~{Boccaletti}, P.~{Bourget}, V.~{Christiaens},
  P.~{Forsberg}, F.~{Gonte}, S.~{Habraken}, C.~{Hanot}, M.~{Karlsson},
  M.~{Kasper}, J.-L. {Lizon}, K.~{Muzic}, R.~{Olivier}, E.~{Pe{\~n}a},
  N.~{Slusarenko}, L.~E. {Tacconi-Garman}, and J.~{Surdej}.
\newblock {L'-band AGPM vector vortex coronagraph's first light on VLT/NACO.
  Discovery of a late-type companion at two beamwidths from an F0V star}.
\newblock \emph{\aap}, 552:\penalty0 L13, April 2013.
\newblock \doi{10.1051/0004-6361/201321315}.

\bibitem[Mawet et~al.(2010)Mawet, Pueyo, Moody, Krist, and
  Serabyn]{doi:10.1117/12.858240}
Dimitri Mawet, Laurent Pueyo, Dwight Moody, John Krist, and Eugene Serabyn.
\newblock The vector vortex coronagraph: sensitivity to central obscuration,
  low-order aberrations, chromaticism, and polarization.
\newblock pages 773914--773914--13, 2010.
\newblock \doi{10.1117/12.858240}.
\newblock URL \url{+ http://dx.doi.org/10.1117/12.858240}.

\bibitem[McBride et~al.(2011)McBride, Graham, Macintosh, Beckwith, Marois,
  Poyneer, and Wiktorowicz]{2011PASP...McBride}
James McBride, James~R. Graham, Bruce Macintosh, Steven V.~W. Beckwith,
  Christian Marois, Lisa~A. Poyneer, and Sloane~J. Wiktorowicz.
\newblock Experimental design for the gemini planet imager.
\newblock \emph{Publications of the Astronomical Society of the Pacific},
  123\penalty0 (904):\penalty0 pp. 692--708, 2011.
\newblock ISSN 00046280.
\newblock URL \url{http://www.jstor.org/stable/10.1086/660733}.

\bibitem[{Metchev} and {Hillenbrand}(2009)]{2009ApJS..181...62M}
S.~A. {Metchev} and L.~A. {Hillenbrand}.
\newblock {The Palomar/Keck Adaptive Optics Survey of Young Solar Analogs:
  Evidence for a Universal Companion Mass Function}.
\newblock \emph{\apjs}, 181:\penalty0 62--109, March 2009.
\newblock \doi{10.1088/0067-0049/181/1/62}.

\bibitem[{Nakajima} et~al.(1995){Nakajima}, {Oppenheimer}, {Kulkarni},
  {Golimowski}, {Matthews}, and {Durrance}]{Nakajima:95}
T.~{Nakajima}, B.R. {Oppenheimer}, S.R. {Kulkarni}, D.A. {Golimowski},
  K.~{Matthews}, and S.T. {Durrance}.
\newblock \emph{Nature}, 378:\penalty0 463, November 1995.
\newblock \doi{10.1038/378463a0}.

\bibitem[{Oppenheimer} et~al.(1997){Oppenheimer}, {Palmer}, {Dekany},
  {Sivaramakrishnan}, {Ealey}, and {Price}]{1997SPIE.3126..569O}
B.~R. {Oppenheimer}, D.~{Palmer}, R.~G. {Dekany}, A.~{Sivaramakrishnan}, M.~A.
  {Ealey}, and T.~R. {Price}.
\newblock {Investigating a Xin{$\xi$}tics Inc. deformable mirror [3126-75]}.
\newblock In {R.~K.~Tyson \& R.~Q.~Fugate}, editor, \emph{Society of
  Photo-Optical Instrumentation Engineers (SPIE) Conference Series}, volume
  3126 of \emph{Society of Photo-Optical Instrumentation Engineers (SPIE)
  Conference Series}, page 569, October 1997.

\bibitem[{Oppenheimer} et~al.(2000){Oppenheimer}, {Dekany}, {Hayward},
  {Brandl}, {Troy}, and {Bloemhof}]{2000SPIE.4007..899O}
B.~R. {Oppenheimer}, R.~G. {Dekany}, T.~L. {Hayward}, B.~{Brandl}, M.~{Troy},
  and E.~E. {Bloemhof}.
\newblock {Companion detection limits with adaptive optics coronagraphy}.
\newblock In {P.~L.~Wizinowich}, editor, \emph{Society of Photo-Optical
  Instrumentation Engineers (SPIE) Conference Series}, volume 4007 of
  \emph{Society of Photo-Optical Instrumentation Engineers (SPIE) Conference
  Series}, pages 899--905, July 2000.

\bibitem[{Oppenheimer} et~al.(2012){Oppenheimer}, {Beichman}, {Brenner},
  {Burruss}, {Cady}, {Crepp}, {Hillenbrand}, {Hinkley}, {Ligon}, {Lockhart},
  {Parry}, {Pueyo}, {Rice}, {Roberts}, {Roberts}, {Shao}, {Sivaramakrishnan},
  {Soummer}, {Vasisht}, {Vescelus}, {Wallace}, {Zhai}, and
  {Zimmerman}]{2012SPIE.8447E..20O}
B.~R. {Oppenheimer}, C.~{Beichman}, D.~{Brenner}, R.~{Burruss}, E.~{Cady},
  J.~{Crepp}, L.~{Hillenbrand}, S.~{Hinkley}, E.~R. {Ligon}, T.~{Lockhart},
  I.~{Parry}, L.~{Pueyo}, E.~{Rice}, L.~C. {Roberts}, J.~{Roberts}, M.~{Shao},
  A.~{Sivaramakrishnan}, R.~{Soummer}, G.~{Vasisht}, F.~{Vescelus}, J.~K.
  {Wallace}, C.~{Zhai}, and N.~{Zimmerman}.
\newblock {Project 1640: the world's first ExAO coronagraphic hyperspectral
  imager for comparative planetary science}.
\newblock In \emph{Society of Photo-Optical Instrumentation Engineers (SPIE)
  Conference Series}, volume 8447 of \emph{Society of Photo-Optical
  Instrumentation Engineers (SPIE) Conference Series}, July 2012.
\newblock \doi{10.1117/12.926419}.

\bibitem[{Oppenheimer} et~al.(2013){Oppenheimer}, {Baranec}, {Beichman},
  {Brenner}, {Burruss}, {Cady}, {Crepp}, {Dekany}, {Fergus}, {Hale},
  {Hillenbrand}, {Hinkley}, {Hogg}, {King}, {Ligon}, {Lockhart}, {Nilsson},
  {Parry}, {Pueyo}, {Rice}, {Roberts}, {Roberts}, {Shao}, {Sivaramakrishnan},
  {Soummer}, {Truong}, {Vasisht}, {Veicht}, {Vescelus}, {Wallace}, {Zhai}, and
  {Zimmerman}]{2013ApJ...768...24O}
B.~R. {Oppenheimer}, C.~{Baranec}, C.~{Beichman}, D.~{Brenner}, R.~{Burruss},
  E.~{Cady}, J.~R. {Crepp}, R.~{Dekany}, R.~{Fergus}, D.~{Hale},
  L.~{Hillenbrand}, S.~{Hinkley}, D.~W. {Hogg}, D.~{King}, E.~R. {Ligon},
  T.~{Lockhart}, R.~{Nilsson}, I.~R. {Parry}, L.~{Pueyo}, E.~{Rice}, J.~E.
  {Roberts}, L.~C. {Roberts}, Jr., M.~{Shao}, A.~{Sivaramakrishnan},
  R.~{Soummer}, T.~{Truong}, G.~{Vasisht}, A.~{Veicht}, F.~{Vescelus}, J.~K.
  {Wallace}, C.~{Zhai}, and N.~{Zimmerman}.
\newblock {Reconnaissance of the HR 8799 Exosolar System. I. Near-infrared
  Spectroscopy}.
\newblock \emph{\apj}, 768:\penalty0 24, May 2013.
\newblock \doi{10.1088/0004-637X/768/1/24}.

\bibitem[{Poyneer} and {Macintosh}(2004)]{2004JOSAA..21..810P}
L.~A. {Poyneer} and B.~{Macintosh}.
\newblock {Spatially filtered wave-front sensor for high-order adaptive
  optics}.
\newblock \emph{Journal of the Optical Society of America A}, 21:\penalty0
  810--819, May 2004.
\newblock \doi{10.1364/JOSAA.21.000810}.

\bibitem[Poyneer et~al.(2003)Poyneer, Troy, Macintosh, and Gavel]{Poyneer:03}
Lisa~A. Poyneer, Mitchell Troy, Bruce Macintosh, and Donald~T. Gavel.
\newblock Experimental validation of fourier-transform wave-front
  reconstruction at the palomar observatory.
\newblock \emph{Opt. Lett.}, 28\penalty0 (10):\penalty0 798--800, May 2003.
\newblock \doi{10.1364/OL.28.000798}.
\newblock URL \url{http://ol.osa.org/abstract.cfm?URI=ol-28-10-798}.

\bibitem[Ren et~al.(2005)Ren, Dekany, and Britton]{Ren:05}
Hongwu Ren, Richard Dekany, and Matthew Britton.
\newblock Large-scale wave-front reconstruction for adaptive optics systems by
  use of a recursive filtering algorithm.
\newblock \emph{Appl. Opt.}, 44\penalty0 (13):\penalty0 2626--2637, May 2005.
\newblock \doi{10.1364/AO.44.002626}.
\newblock URL \url{http://ao.osa.org/abstract.cfm?URI=ao-44-13-2626}.

\bibitem[{Roberts} et~al.(2010){Roberts}, {Bouchez}, {Burruss}, {Dekany},
  {Guiwits}, and {Troy}]{2010SPIE.7736E..81R}
J.~{Roberts}, A.~H. {Bouchez}, R.~S. {Burruss}, R.~G. {Dekany}, S.~R.
  {Guiwits}, and M.~{Troy}.
\newblock {Optical characterization of the PALM-3000 3388-actuator deformable
  mirror}.
\newblock In \emph{Society of Photo-Optical Instrumentation Engineers (SPIE)
  Conference Series}, volume 7736 of \emph{Society of Photo-Optical
  Instrumentation Engineers (SPIE) Conference Series}, July 2010.
\newblock \doi{10.1117/12.857815}.

\bibitem[{Roberts} et~al.(2008){Roberts}, {Bouchez}, {Angione}, {Burruss},
  {Cromer}, {Dekany}, {Guiwits}, {Henning}, {Hickey}, {Kibblewhite}, {McKenna},
  {Moore}, {Petrie}, {Shelton}, {Thicksten}, {Trinh}, {Tripathi}, {Troy},
  {Truong}, and {Velur}]{2008SPIE.7015E..74R}
J.~E. {Roberts}, A.~H. {Bouchez}, J.~{Angione}, R.~S. {Burruss}, J.~L.
  {Cromer}, R.~G. {Dekany}, S.~R. {Guiwits}, J.~R. {Henning}, J.~{Hickey},
  E.~{Kibblewhite}, D.~L. {McKenna}, A.~M. {Moore}, H.~L. {Petrie}, J.~C.
  {Shelton}, R.~P. {Thicksten}, T.~{Trinh}, R.~{Tripathi}, M.~{Troy},
  T.~{Truong}, and V.~{Velur}.
\newblock {Facilitizing the Palomar AO laser guide star system}.
\newblock In \emph{Society of Photo-Optical Instrumentation Engineers (SPIE)
  Conference Series}, volume 7015 of \emph{Society of Photo-Optical
  Instrumentation Engineers (SPIE) Conference Series}, July 2008.
\newblock \doi{10.1117/12.790042}.

\bibitem[{Roberts} et~al.(2012){Roberts}, {Dekany}, {Burruss}, {Baranec},
  {Bouchez}, {Croner}, {Guiwits}, {Hale}, {Henning}, {Palmer}, {Troy},
  {Truong}, and {Zolkower}]{2012SPIE.8447E..0YR}
J.~E. {Roberts}, R.~G. {Dekany}, R.~S. {Burruss}, C.~{Baranec}, A.~{Bouchez},
  E.~E. {Croner}, S.~R. {Guiwits}, D.~D.~S. {Hale}, J.~R. {Henning}, D.~L.
  {Palmer}, M.~{Troy}, T.~N. {Truong}, and J.~{Zolkower}.
\newblock {Results from the PALM-3000 high-order adaptive optics system}.
\newblock In \emph{Society of Photo-Optical Instrumentation Engineers (SPIE)
  Conference Series}, volume 8447 of \emph{Society of Photo-Optical
  Instrumentation Engineers (SPIE) Conference Series}, July 2012.
\newblock \doi{10.1117/12.926477}.

\bibitem[{Roberts} et~al.(2004){Roberts}, {Perrin}, {Marchis},
  {Sivaramakrishnan}, {Makidon}, {Christou}, {Macintosh}, {Poyneer}, {van Dam},
  and {Troy}]{2004SPIE.5490..504R}
L.~C. {Roberts}, Jr., M.~D. {Perrin}, F.~{Marchis}, A.~{Sivaramakrishnan},
  R.~B. {Makidon}, J.~C. {Christou}, B.~A. {Macintosh}, L.~A. {Poyneer}, M.~A.
  {van Dam}, and M.~{Troy}.
\newblock {Is that really your Strehl ratio?}
\newblock In D.~{Bonaccini Calia}, B.~L. {Ellerbroek}, and R.~{Ragazzoni},
  editors, \emph{Society of Photo-Optical Instrumentation Engineers (SPIE)
  Conference Series}, volume 5490 of \emph{Society of Photo-Optical
  Instrumentation Engineers (SPIE) Conference Series}, pages 504--515, October
  2004.
\newblock \doi{10.1117/12.549115}.

\bibitem[{Rousset} et~al.(2003){Rousset}, {Lacombe}, {Puget}, {Hubin},
  {Gendron}, {Fusco}, {Arsenault}, {Charton}, {Feautrier}, {Gigan}, {Kern},
  {Lagrange}, {Madec}, {Mouillet}, {Rabaud}, {Rabou}, {Stadler}, and
  {Zins}]{2003SPIE.4839..140R}
G.~{Rousset}, F.~{Lacombe}, P.~{Puget}, N.~N. {Hubin}, E.~{Gendron},
  T.~{Fusco}, R.~{Arsenault}, J.~{Charton}, P.~{Feautrier}, P.~{Gigan}, P.~Y.
  {Kern}, A.-M. {Lagrange}, P.-Y. {Madec}, D.~{Mouillet}, D.~{Rabaud},
  P.~{Rabou}, E.~{Stadler}, and G.~{Zins}.
\newblock {NAOS, the first AO system of the VLT: on-sky performance}.
\newblock In {P.~L.~Wizinowich \& D.~Bonaccini}, editor, \emph{Society of
  Photo-Optical Instrumentation Engineers (SPIE) Conference Series}, volume
  4839 of \emph{Society of Photo-Optical Instrumentation Engineers (SPIE)
  Conference Series}, pages 140--149, February 2003.
\newblock \doi{10.1117/12.459332}.

\bibitem[{Sch{\"o}del} et~al.(2013){Sch{\"o}del}, {Yelda}, {Ghez}, {Girard},
  {Labadie}, {Rebolo}, {P{\'e}rez-Garrido}, and {Morris}]{2013MNRAS.429.1367S}
R.~{Sch{\"o}del}, S.~{Yelda}, A.~{Ghez}, J.~H. {Girard}, L.~{Labadie},
  R.~{Rebolo}, A.~{P{\'e}rez-Garrido}, and M.~R. {Morris}.
\newblock {Holographic imaging of crowded fields: high angular resolution
  imaging with excellent quality at very low cost}.
\newblock \emph{\mnras}, 429:\penalty0 1367--1375, February 2013.
\newblock \doi{10.1093/mnras/sts420}.

\bibitem[Serabyn et~al.(2009)Serabyn, Mawet, Bloemhof, Haguenauer, Mennesson,
  Wallace, and Hickey]{Serabyn:09}
E.~Serabyn, D.~Mawet, E.~Bloemhof, P.~Haguenauer, B.~Mennesson, K.~Wallace, and
  J.~Hickey.
\newblock Imaging faint brown dwarf companions close to bright stars with a
  small, well-corrected telescope aperture.
\newblock \emph{The Astrophysical Journal}, 696\penalty0 (1):\penalty0 40,
  2009.
\newblock URL \url{http://stacks.iop.org/0004-637X/696/i=1/a=40}.

\bibitem[{Shi} et~al.(2003){Shi}, {MacMartin}, {Troy}, {Brack}, {Burruss}, and
  {Dekany}]{2003SPIE.4839.1035S}
F.~{Shi}, D.~G. {MacMartin}, M.~{Troy}, G.~L. {Brack}, R.~S. {Burruss}, and
  R.~G. {Dekany}.
\newblock {Sparse matrix wavefront reconstruction: simulations and
  experiments}.
\newblock In {P.~L.~Wizinowich \& D.~Bonaccini}, editor, \emph{Society of
  Photo-Optical Instrumentation Engineers (SPIE) Conference Series}, volume
  4839 of \emph{Society of Photo-Optical Instrumentation Engineers (SPIE)
  Conference Series}, pages 1035--1044, February 2003.
\newblock \doi{10.1117/12.457134}.

\bibitem[{Swartzlander, Jr.} et~al.(2008){Swartzlander, Jr.}, {Ford},
  {Abdul-Malik}, {Close}, {Peters}, {Palacios}, and {Wilson}]{Swartzlander08}
G.~A. {Swartzlander, Jr.}, E.~L. {Ford}, R.~S. {Abdul-Malik}, L.~M. {Close},
  M.~A. {Peters}, D.~M. {Palacios}, and D.~W. {Wilson}.
\newblock Astronomical demonstration of anoptical vortex coronagraph.
\newblock \emph{Opt. Express}, 16\penalty0 (14):\penalty0 10200--10207, Jul
  2008.
\newblock \doi{10.1364/OE.16.010200}.
\newblock URL
  \url{http://www.opticsexpress.org/abstract.cfm?URI=oe-16-14-10200}.

\bibitem[Tecza et~al.(2012)Tecza, Thatte, Clarke, Lynn, Freeman, Roberts, and
  Dekany]{doi:10.1117/12.925328}
Matthias Tecza, Niranjan Thatte, Fraser Clarke, James Lynn, David Freeman,
  Jennifer Roberts, and Richard Dekany.
\newblock 15x optical zoom and extreme optical image stabilisation: diffraction
  limited integral field spectroscopy with the oxford swift spectrograph.
\newblock pages 844622--844622--8, 2012.
\newblock \doi{10.1117/12.925328}.
\newblock URL \url{+ http://dx.doi.org/10.1117/12.925328}.

\bibitem[{Thatte} et~al.(2010){Thatte}, {Tecza}, {Clarke}, {Goodsall},
  {Fogarty}, {Houghton}, {Salter}, {Scott}, {Davies}, {Bouchez}, and
  {Dekany}]{2010SPIE.7735E.258T}
N.~{Thatte}, M.~{Tecza}, F.~{Clarke}, T.~{Goodsall}, L.~{Fogarty},
  R.~{Houghton}, G.~{Salter}, N.~{Scott}, R.~L. {Davies}, A.~{Bouchez}, and
  R.~{Dekany}.
\newblock {The Oxford SWIFT Spectrograph: first commissioning and on-sky
  results}.
\newblock In \emph{Society of Photo-Optical Instrumentation Engineers (SPIE)
  Conference Series}, volume 7735 of \emph{Society of Photo-Optical
  Instrumentation Engineers (SPIE) Conference Series}, July 2010.
\newblock \doi{10.1117/12.857484}.

\bibitem[{Troy} et~al.(2000){Troy}, {Dekany}, {Brack}, {Oppenheimer},
  {Bloemhof}, {Trinh}, {Dekens}, {Shi}, {Hayward}, and
  {Brandl}]{2000SPIE.4007...31T}
M.~{Troy}, R.~G. {Dekany}, G.~{Brack}, B.~R. {Oppenheimer}, E.~E. {Bloemhof},
  T.~{Trinh}, F.~G. {Dekens}, F.~{Shi}, T.~L. {Hayward}, and B.~{Brandl}.
\newblock {Palomar adaptive optics project: status and performance}.
\newblock In {P.~L.~Wizinowich}, editor, \emph{Society of Photo-Optical
  Instrumentation Engineers (SPIE) Conference Series}, volume 4007 of
  \emph{Society of Photo-Optical Instrumentation Engineers (SPIE) Conference
  Series}, pages 31--40, July 2000.

\bibitem[{Truong} et~al.(2003){Truong}, {Brack}, {Troy}, {Trinh}, {Shi}, and
  {Dekany}]{2003SPIE.4839..911T}
T.~{Truong}, G.~L. {Brack}, M.~{Troy}, T.~{Trinh}, F.~{Shi}, and R.~G.
  {Dekany}.
\newblock {Real-time wavefront processors for the next generation of adaptive
  optics systems: a design and analysis}.
\newblock In {P.~L.~Wizinowich \& D.~Bonaccini}, editor, \emph{Society of
  Photo-Optical Instrumentation Engineers (SPIE) Conference Series}, volume
  4839 of \emph{Society of Photo-Optical Instrumentation Engineers (SPIE)
  Conference Series}, pages 911--922, February 2003.
\newblock \doi{10.1117/12.459353}.

\bibitem[{Truong} et~al.(2008){Truong}, {Bouchez}, {Dekany}, {Shelton}, {Troy},
  {Angione}, {Burruss}, {Cromer}, {Guiwits}, and
  {Roberts}]{2008SPIE.7015E..95T}
T.~N. {Truong}, A.~H. {Bouchez}, R.~G. {Dekany}, J.~C. {Shelton}, M.~{Troy},
  J.~R. {Angione}, R.~S. {Burruss}, J.~L. {Cromer}, S.~R. {Guiwits}, and J.~E.
  {Roberts}.
\newblock {Real-time wavefront control for the PALM-3000 high order adaptive
  optics system}.
\newblock In \emph{Society of Photo-Optical Instrumentation Engineers (SPIE)
  Conference Series}, volume 7015 of \emph{Society of Photo-Optical
  Instrumentation Engineers (SPIE) Conference Series}, July 2008.
\newblock \doi{10.1117/12.790457}.

\bibitem[Vasisht(2013)]{Vasisht:13}
Gautam Vasisht.
\newblock personal communication, mar 2013.

\bibitem[{Velur} et~al.(2004){Velur}, {Kibblewhite}, {Dekany}, {Troy},
  {Petrie}, {Thicksten}, {Brack}, {Trin}, and {Cheselka}]{2004SPIE.5490.1033V}
V.~{Velur}, E.~J. {Kibblewhite}, R.~G. {Dekany}, M.~{Troy}, H.~L. {Petrie},
  R.~P. {Thicksten}, G.~{Brack}, T.~{Trin}, and M.~{Cheselka}.
\newblock {Implementation of the Chicago sum frequency laser at Palomar laser
  guide star test bed}.
\newblock In {D.~Bonaccini Calia, B.~L.~Ellerbroek, \& R.~Ragazzoni}, editor,
  \emph{Society of Photo-Optical Instrumentation Engineers (SPIE) Conference
  Series}, volume 5490 of \emph{Society of Photo-Optical Instrumentation
  Engineers (SPIE) Conference Series}, pages 1033--1040, October 2004.
\newblock \doi{10.1117/12.550675}.

\bibitem[{Velur} et~al.(2006){Velur}, {Flicker}, {Platt}, {Britton}, {Dekany},
  {Troy}, {Roberts}, {Shelton}, and {Hickey}]{2006SPIE.6272E.169V}
V.~{Velur}, R.~C. {Flicker}, B.~C. {Platt}, M.~C. {Britton}, R.~G. {Dekany},
  M.~{Troy}, J.~E. {Roberts}, J.~C. {Shelton}, and J.~{Hickey}.
\newblock {Multiple guide star tomography demonstration at Palomar
  observatory}.
\newblock In \emph{Society of Photo-Optical Instrumentation Engineers (SPIE)
  Conference Series}, volume 6272 of \emph{Society of Photo-Optical
  Instrumentation Engineers (SPIE) Conference Series}, July 2006.
\newblock \doi{10.1117/12.671666}.

\bibitem[{Wyatt}(2008)]{2008ARA&A..46..339W}
M.~C. {Wyatt}.
\newblock {Evolution of Debris Disks}.
\newblock \emph{\araa}, 46:\penalty0 339--383, September 2008.
\newblock \doi{10.1146/annurev.astro.45.051806.110525}.

\bibitem[{Wyatt} et~al.(1999){Wyatt}, {Dermott}, {Telesco}, {Fisher}, {Grogan},
  {Holmes}, and {Pi{\~n}a}]{1999ApJ...527..918W}
M.~C. {Wyatt}, S.~F. {Dermott}, C.~M. {Telesco}, R.~S. {Fisher}, K.~{Grogan},
  E.~K. {Holmes}, and R.~K. {Pi{\~n}a}.
\newblock {How Observations of Circumstellar Disk Asymmetries Can Reveal Hidden
  Planets: Pericenter Glow and Its Application to the HR 4796 Disk}.
\newblock \emph{\apj}, 527:\penalty0 918--944, December 1999.
\newblock \doi{10.1086/308093}.

\bibitem[{Zimmerman} et~al.(2011){Zimmerman}, {Brenner}, {Oppenheimer},
  {Parry}, {Hinkley}, {Hunt}, and {Roberts}]{2011PASP..123..746Z}
N.~{Zimmerman}, D.~{Brenner}, B.~R. {Oppenheimer}, I.~R. {Parry}, S.~{Hinkley},
  S.~{Hunt}, and R.~{Roberts}.
\newblock {A Data-Cube Extraction Pipeline for a Coronagraphic Integral Field
  Spectrograph}.
\newblock \emph{\pasp}, 123:\penalty0 746--763, June 2011.
\newblock \doi{10.1086/660818}.

\end{thebibliography}

\end{document}